\documentclass[aps,prd,nofootinbib,twocolumn,superscriptaddress,preprintnumbers,letterpaper,natbib,nofootinbib]{revtex4-1}

\pdfoutput=1
\usepackage{amssymb,amsmath,latexsym,mathrsfs}
\usepackage[normalem]{ulem}
\usepackage{url}
\usepackage{epsfig,graphicx,verbatim,xspace,multirow,mathtools}
\usepackage{enumitem}
\usepackage{graphicx}
\usepackage[usenames,dvipsnames]{color}
\usepackage[breaklinks,colorlinks,urlcolor=blue,citecolor=blue,linkcolor=magenta]{hyperref}
\usepackage{comment}
\usepackage{bm}
\usepackage[table]{xcolor}
\usepackage{ulem}

\usepackage{hhline}
\usepackage{multirow}

\makeatletter

\begin{document}

\title{Primordial lepton asymmetries in the precision cosmology era: \\ Current status and future sensitivities from BBN and the CMB}

\preprint{TUM-HEP-1413/22}
\author{Miguel Escudero}
\email{miguel.escudero@tum.de}
\thanks{ORCID: \href{https://orcid.org/0000-0002-4487-8742}{0000-0002-4487-8742}}
\affiliation{Physik-Department, Technische Universit{\"{a}}t, M{\"{u}}nchen, James-Franck-Stra{\ss}e, 85748 Garching, Germany}

\author{Alejandro Ibarra}
\email{ibarra@tum.de}
\thanks{ORCID: \href{https://orcid.org/0000-0001-9935-5247}{0000-0001-9935-5247}}
\affiliation{Physik-Department, Technische Universit{\"{a}}t, M{\"{u}}nchen, James-Franck-Stra{\ss}e, 85748 Garching, Germany}

\author{Victor Maura}
\email{victor.maura.breick@tum.de}
\thanks{ORCID: \href{https://orcid.org/0000-0003-2930-6356}{0000-0003-2930-6356}}
\affiliation{Physik-Department, Technische Universit{\"{a}}t, M{\"{u}}nchen, James-Franck-Stra{\ss}e, 85748 Garching, Germany}

\begin{abstract}
\noindent Using a new sample of extremely metal poor systems, the EMPRESS survey has recently reported a primordial helium abundance that is $3\sigma$ smaller than the prediction from the standard big bang nucleosynthesis (BBN) scenario. This measurement could be interpreted as a hint for a primordial lepton asymmetry in the electron neutrino flavor. Motivated by the EMPRESS results, we present a comprehensive analysis of the lepton asymmetry using measurements of the abundances of primordial elements, along with cosmic microwave background (CMB) data from Planck. Assuming that there is no dark radiation in our Universe, we find an electron neutrino chemical potential $\xi_{\nu_e} = 0.043 \pm 0.015$, which deviates from zero by $2.9\sigma$. If no assumption is made on the abundance of dark radiation in the Universe, the chemical potential is  $\xi_{\nu_e} = 0.046 \pm 0.021$, which deviates from zero by $2.2\sigma$. We also find that this result is rather insensitive to the choice of nuclear reaction rates. If the true helium abundance corresponds to the EMPRESS central value, future CMB observations from the Simons Observatory and CMB-S4 will increase the significance for a nonzero lepton asymmetry to $4\sigma$ and $5\sigma$ respectively, assuming no dark radiation, or to $3\sigma$ when no assumption is made on the abundance of dark radiation.  
\end{abstract}

\maketitle
{
   \hypersetup{linkcolor=black}
   \setlength\parskip{-0.2pt}
   \setlength\parindent{0.0pt}
}

\section{Introduction}\label{sec:intro}

 We appear to be living in a Universe composed mostly by matter and with very little antimatter~\cite{Kolb:1990vq}. This strongly suggests the existence of a mechanism generating a primordial asymmetry between baryons and antibaryons in the very early Universe~\cite{Gorbunov:2011zz}. The abundance of baryons in the Universe has now been measured with $\lesssim 1\%$ precision using observations of the cosmic microwave background (CMB)~\cite{planck}, and by comparing the observed and predicted primordial element abundances as synthesized during big bang nucleosynthesis (BBN)~\cite{Mossa:2020gjc,Pisanti:2020efz,Pitrou:2020etk,Yeh:2022heq}. These observations point to a baryon asymmetry, defined as the number density of baryons minus antibaryons normalized to the photon number density, given by $\eta_B \equiv (n_B-n_{\overline{B}})/n_\gamma  =  (6.14 \pm 0.04)\times 10^{-10}$~\cite{planck}.

However, much less is known about the primordial lepton asymmetries, $\eta_{L_\alpha}$, with $\alpha = e,\,\mu,\,\tau$. Naively one would expect the lepton and baryon asymmetries to be of similar magnitude, due to sphaleron transitions in the early Universe~\cite{Kuzmin:1985mm,Khlebnikov:1988sr,Harvey:1990qw,Dreiner:1992vm}. However, this does not necessarily need to be the case. Indeed, several scenarios have been constructed where the lepton asymmetries at the time of BBN can be much larger than the baryon asymmetry. In these scenarios the lepton asymmetry is typically generated at temperatures below the sphaleron freeze-out via Affleck-Dine leptogenesis~\cite{Casas:1997gx,Dolgov:1989us}, decays of topological defects~\cite{Bajc:1997ky}, freeze-in leptogenesis~\cite{Asaka:2005pn,Asaka:2005an}, resonant-leptogenesis~\cite{Pilaftsis:2003gt,Borah:2022uos} or Q-ball decays~\cite{Kawasaki:2002hq,Kawasaki:2022hvx}. Furthermore, there are scenarios where large lepton asymmetries are generated before sphaleron freeze-out but in which the total lepton asymmetry in the Universe is zero~\cite{March-Russell:1999hpw}, see also~\cite{Domcke:2022uue} for new further cosmological constraints on such scenarios.

The main effect of a nonzero electron lepton asymmetry at the time of BBN is to change the value of the primordial helium abundance, $Y_{\rm P}$~\cite{Sarkar:1995dd,Iocco:2008va,Pitrou:2018cgg,Lesgourgues:2013sjj,Serpico:2005bc,Mangano:2011ip,Chu:2006ua,Simha:2008mt}. This happens because electron neutrinos participate in processes that interconvert protons and neutrons, such as the weak interaction process $ n \,\nu_e \leftrightarrow p \,e^-$. At the time of BBN, corresponding to $T_\gamma \simeq 0.073\,{\rm MeV}$~\cite{Mukhanov:2003xs}, almost all of the neutrons present in the plasma form ${}^4{\rm He}$. Therefore, any excess of $\nu_e$ over $\bar{\nu}_e$ in the early Universe will translate into a smaller abundance of neutrons, and correspondingly to a smaller helium abundance compared to the Standard Model expectation. 

\begin{figure}[t]
    \centering
    \includegraphics[width=0.5\textwidth]{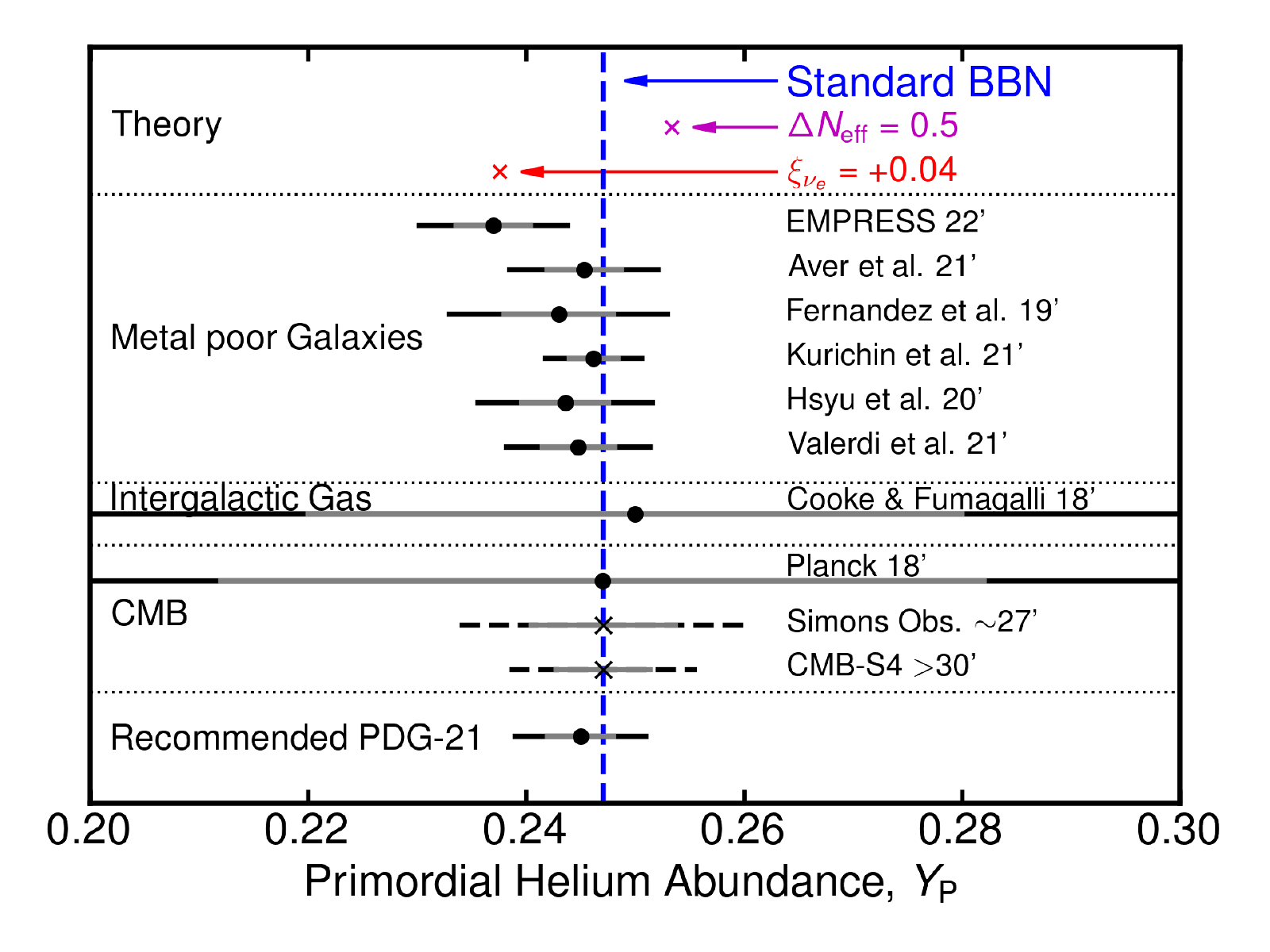}
    \vspace{-0.6cm}
    \caption{Current status in the measurements and the theoretical determinations of the primordial helium abundance, and forecasts for the upcoming Simons Observatory and CMB-S4.}
    \label{fig:HeliumStatus}
\end{figure}

The most common method to determine the primordial helium abundance consists in measuring the helium abundance in metal poor galaxies, and extrapolating the value to zero metallicity~\cite{Aver:2020fon,Fernandez:2019hds,2020ApJ...896...77H,2021MNRAS.505.3624V,Kurichin:2021ppm}. Alternatively, the helium abundance could be measured in intergalactic gas clouds~\cite{Cooke:2018qzw}. In a cosmological context, the helium abundance at the time of recombination affects the number of free electrons, thus leaving an imprint in the CMB temperature and polarization power spectra at small angular scales~\cite{Planck:2018nkj,planck_legacy}. A summary of recent determinations is shown in Fig. ~\ref{fig:HeliumStatus}, and show a fairly good agreement with the Standard Model expectations. On the other hand, very recently the EMPRESS survey~\cite{Matsumoto:2022tlr} increased the sample of extremely metal poor systems, and reported a value for the primordial helium abundance which is $3\sigma$ smaller than the value predicted by the Standard Model~\cite{Pitrou:2018cgg}, suggesting the existence of a nonzero (electron) lepton asymmetry. 

Motivated by the recent result by the EMPRESS survey, we will undertake a comprehensive study of current BBN and CMB constraints on the lepton asymmetries (see \cite{Oldengott:2017tzj,Pitrou:2018cgg,Seto:2021tad,Kumar:2022vee,Matsumoto:2022tlr,Burns:2022hkq} for related analyses). Moreover, we will also explore the sensitivity of upcoming cosmological observations to a nonzero primordial lepton asymmetry. Specifically, we will make a forecast for the upcoming Simons Observatory~\cite{SimonsObservatory:2018koc,SimonsObservatory:2019qwx}, a fully funded ground base experiment that is expected to finalize data taking by 2027, and for  a Stage-IV experiment such as CMB-S4~\cite{CMB-S4:2016ple,Abazajian:2019eic}, which will provide a measurement of the primordial helium abundance with a precision of $\sim 1\%-2\%$ (see Fig. ~\ref{fig:HeliumStatus}).

This work is structured as follows. In Sec.~\ref{sec:Implications} we briefly review the impact of a nonzero lepton asymmetry for BBN and CMB observations. In Sec.~\ref{sec:results_today}, we summarize the current information on the abundance of primordial elements and we present the result of our analysis of the lepton asymmetry. Then, in  Sec.~\ref{sec:forecast} we present forecasts for the Simons Observatory and CMB-S4. Lastly, in Sec.~\ref{sec:conclusions} we present our conclusions.

\section{Implications of a Primordial Lepton Asymmetry for BBN and the CMB}\label{sec:Implications}

The primordial lepton asymmetry is normally parametrized by the (comoving) neutrino chemical potential, $\xi_\nu$, through~\cite{Iocco:2008va}:
\begin{align}\label{eq:numberasymmetry}
    \eta_{L_\alpha} &\equiv \frac{n_{\nu_\alpha}-n_{\bar{\nu}_\alpha}}{n_\gamma} = \frac{1}{12\zeta(3)} \left[\frac{T_{\nu_\alpha}}{T_\gamma}\right]^{3} \left(\pi^2 \xi_{\nu_\alpha} + \xi_{\nu_\alpha}^3\right) \,, \\
    &\simeq 0.25 \, \xi_{\nu_\alpha} \left[  1 + \xi_{\nu_\alpha}^2/\pi^2 \right] \,,\nonumber
\end{align}
where $\zeta(3) \simeq 1.20206$, and where in the last step we have used the value of $T_\gamma/T_\nu$  expected from neutrino decoupling in the Standard Model~\cite{EscuderoAbenza:2020cmq}. 

The implications of a nonzero lepton asymmetry in BBN and the CMB have been studied in the past (for reviews, see {\it e.g.}~\cite{Sarkar:1995dd,Iocco:2008va,Pitrou:2018cgg,Lesgourgues:2013sjj}). The effect of a lepton asymmetry in cosmology depends critically upon its flavor. As discussed in the introduction, a nonzero asymmetry in the electron-neutrino flavor alters the helium abundance by changing the rate of proton-to-neutron conversions in the early Universe. More concretely, it leads to a shift in the primordial helium abundance of~\cite{Pitrou:2018cgg}:
\begin{align}\label{eq:Ypvariation}
   Y_{\rm P}(\xi_{\nu_e}) \simeq Y_{\rm P}|_{\rm SBBN} \times e^{-0.96\,\xi_{\nu_e}}\,,
\end{align}
where $Y_{\rm P}|_{\rm SBBN}$ refers to the primordial helium abundance in the standard BBN scenario, namely when the neutrino chemical potential vanishes, $Y_{\rm P}|_{\rm SBBN} = 0.24709 \pm 0.00017$~\cite{Pitrou:2018cgg}. A nonzero lepton asymmetry also affects the abundances of the rest of the light elements. For deuterium the effect is~\cite{Pitrou:2018cgg}: 
\begin{align}\label{eq:DHvariation}
   {\rm D/H}|_{\rm P}(\xi_{\nu_e})  \simeq {\rm D/H}_{\rm P}|_{\rm SBBN}\times e^{-0.53\,\xi_{\nu_e}} \, \,.
\end{align}
where again, ${\rm D/H}_{\rm P}|_{\rm SBBN}$ refers to the value of the primordial deuterium abundance for a zero lepton asymmetry. It is important to note, however, that in contrast to helium, this abundance is strongly sensitive to the baryon energy density, ${\rm D/H}|_{\rm P} \propto (\Omega_b h^2)^{-1.6}$~\cite{Fields:2019pfx}. Therefore, the sensitivity to $\xi_{\nu_e}$ from ${\rm D/H}|_{\rm P}$ is lost unless $\Omega_bh^2$ is given as an input by other methods.

In addition, the presence of a nonzero asymmetry alters the energy density carried out by neutrinos. It is important to stress that this effect is independent of the flavor of the asymmetry or its sign. This explicitly amounts to a contribution to the number of effective relativistic neutrino species of:
\begin{align}\label{eq:DeltaNeff}
    \Delta N_{\rm eff} = \sum_\alpha^{e,\,\mu,\,\tau}  \left[\frac{30}{7}\left(\frac{\xi_\alpha}{\pi}\right)^2 +\frac{15}{7}\left(\frac{\xi_\alpha}{\pi}\right)^4\right]\,,
\end{align}
where $\Delta N_{\rm eff} \equiv N_{\rm eff} - N_{\rm eff}^{\rm SM} $ with $N_{\rm eff}^{\rm SM} = 3.044(1)$~\cite{EscuderoAbenza:2020cmq,Akita:2020szl,Froustey:2020mcq,Bennett:2020zkv}. Due to neutrino oscillations in the early Universe, one expects $|\xi_{\nu_e}| \simeq |\xi_{\nu_\mu}| \simeq |\xi_{\nu_\tau}|$~\cite{Dolgov:2002ab,Wong:2002fa,Abazajian:2002qx,Froustey:2021azz}. Therefore, and in view of the current constraints on the electron lepton asymmetry $|\xi_{\nu_e}| \lesssim 0.1$, the modification on $\Delta N_{\rm eff}$ due to a nonzero chemical potential is expected to be $\Delta N_{\rm eff}\lesssim 0.01$, much smaller than the current sensitivity of experiments. In what follows we will therefore focus only on the impact of the nonzero lepton asymmetry on $Y_{\rm P}$.

\section{Current constraints on the lepton asymmetries from BBN and CMB data}\label{sec:results_today}

We will analyze the electron neutrino chemical potential from the BBN and CMB data for two possible cosmological scenarios, namely when $N_{\rm eff}=N_{\rm eff}^{\rm SM}=3.044$ or when $N_{\rm eff}$ differs from the SM expectation (corresponding respectively to scenarios without or with dark radiation).

In our analysis we will mainly focus on the implications of the recent helium measurement by EMPRESS~\cite{Matsumoto:2022tlr}:
\begin{align}\label{eq:YPEMPRESS}
Y_{\rm P}|_{\rm EMPRESS} = 0.2370^{+0.0034}_{-0.0033} \,.
\end{align}
which is $3.0\sigma$ lower than the standard BBN prediction. However, we will also consider for comparison the recommended PDG-21 value~\cite{pdg}:
\begin{align}\label{eq:YPPDG}
Y_{\rm P}|_{\rm PDG-21} = 0.245 \pm 0.003 \,.
\end{align}
We will also include the measurement of the  primordial deuterium abundance, which is typically used to constrain the baryon energy density. 
The PDG recommended value reads~\cite{pdg}:  
\begin{align}\label{eq:DHPDG}
{\rm D/H}_{\rm P}|_{\rm PDG-21} = (2.547 \pm 0.025)\times 10^{-5}\,,
\end{align}
which is largely based on the analysis of~\cite{Cooke:2017cwo}.

\begin{figure*}[t]
    \centering
    \begin{tabular}{cc}
      \includegraphics[width=0.5\textwidth]{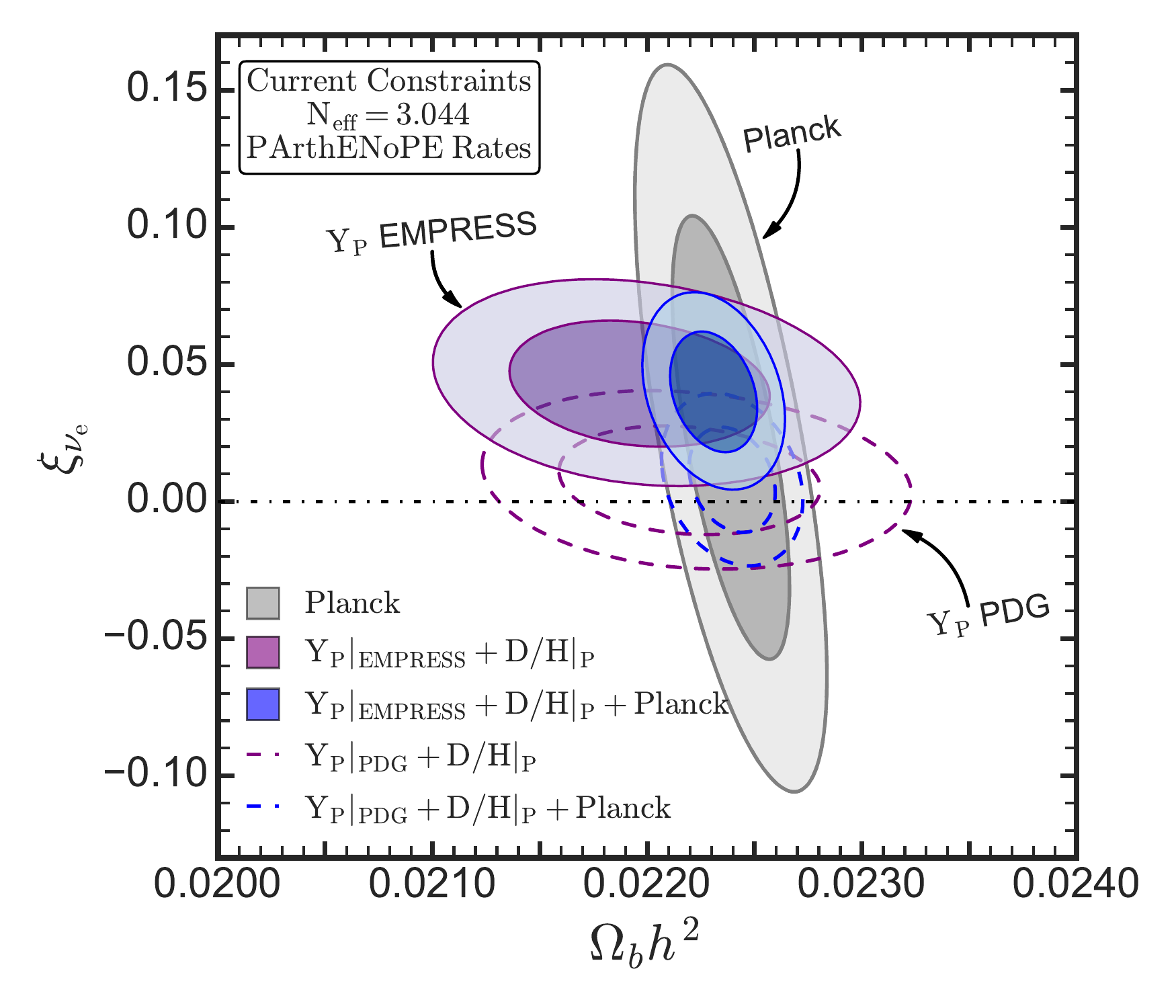}
      &
      \includegraphics[width=0.5\textwidth]{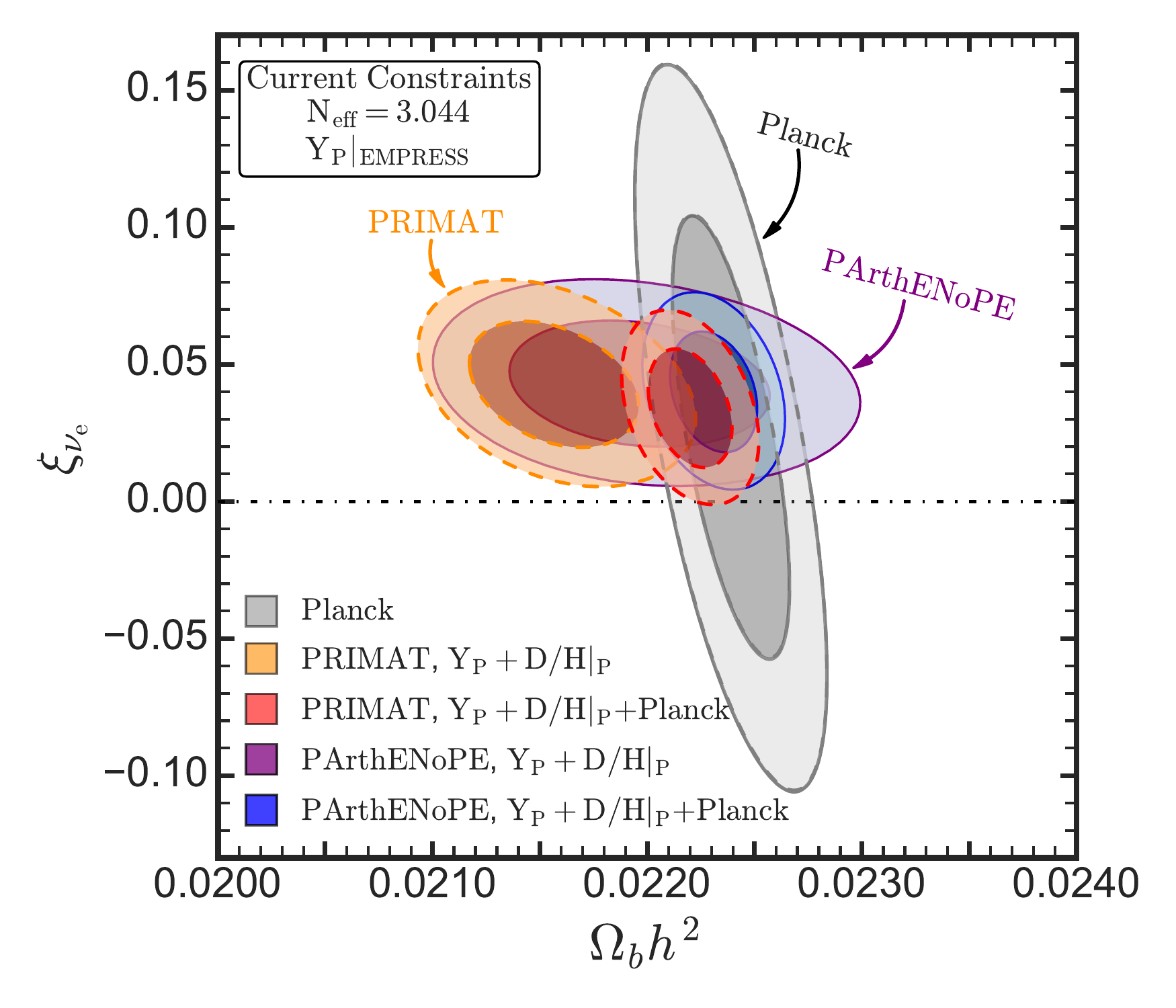}
    \end{tabular}
    \caption{$1$ and $2\sigma$ C.L. regions for $\xi_{\nu_e}$ and $\Omega_bh^2$ from nucleosynthesis data, CMB data, and their combination for a cosmological scenario without dark radiation ({\it i.e.} assuming $N_{\rm eff} = N_{\rm eff}^{\rm SM} =  3.044$). The left panel compares the favored regions for two determinations of the helium abundance (EMPRESS survey and the PDG-21 recommended value) adopting the PArthENoPE nuclear rates, while the right panel compares the favored regions for two choices of the nuclear reaction rates (PArthENoPE or PRIMAT) adopting the EMPRESS measurement of the helium abundance.}
    \label{fig:xi_vs_omega_Nefffix}
\end{figure*}

Lastly, we will also use results from Planck CMB observations~\cite{Planck:2018nkj}, which provide independent determinations of $\Omega_bh^2$, $Y_{\rm P}$ and $N_{\rm eff}$. Concretely, assuming the standard cosmological model, the Planck collaboration reports a baryon energy density
\begin{align}\label{eq:omegabprior}
\Omega_b h^2|_{\rm Planck}  = 0.02242 \pm 0.00014\,,
\end{align}
from combining the full temperature and polarization data, together with CMB lensing and baryon acoustic oscillations.

The Planck collaboration has also made an analysis of the CMB data under the assumption that  $N_{\rm eff} = N_{\rm eff}^{\rm SM}$ but allowing for a non-standard primordial helium abundance. The determination of $Y_{\rm P}$ is correlated with $\Omega_b h^2$ and reads~\cite{planck_legacy}:
\begin{subequations}\label{eq:YPPlanck}
\begin{align}
\Omega_b h^2|_{\rm Planck}  &=  0.02239 \pm 0.00018 \,,\\
Y_{\rm P}|_{\rm Planck}  &= 0.242 \pm 0.012\,, \\
\rho(\Omega_b h^2,\,Y_{\rm P})  &= 0.663\,,
\end{align}
\end{subequations}
where $\rho$ represents the correlation coefficient. Lastly, the Planck collaboration has analyzed the CMB data allowing also for variations in $N_{\rm eff}$. For this scenario, the determination of $\Omega_b h^2$, $Y_{\rm P}$ and $N_{\rm eff}$ reads:
\begin{subequations}\label{eq:YPPlanck_Neff}
\begin{align}
&{\rm Planck}\nonumber \\
\Omega_b h^2|_{\rm Planck}  &= 0.02238 \pm 0.00019 \,,\\
Y_{\rm P}|_{\rm Planck}           &= 0.245 \pm 0.018 \,,\\
N_{\rm eff}                 &= 2.97 \pm 0.29 \,,\\
\rho(\Omega_b h^2,\,Y_{\rm P})   &= +0.273\,,\\
\rho(\Omega_b h^2,\,N_{\rm eff}) &= +0.270\,,\\
\rho(N_{\rm eff},\,Y_{\rm P})  &= -0.686\,.
\end{align}
\end{subequations}
with their corresponding correlation coefficients.

To calculate the abundances of the primordial elements we use the public code {\tt PArthENoPE-v3.0}~\cite{Pisanti:2007hk,Consiglio:2017pot,Gariazzo:2021iiu}. This code takes into account all nuclear reaction rates and weak processes relevant for the nucleosynthesis process in the presence of a primordial lepton asymmetry. At present, there is agreement between all the outputs of this code and the codes used by the other leading groups performing global BBN analyses~\cite{Fields:2019pfx,Pitrou:2018cgg}, with the exception of the primordial deuterium abundance. After the measurement by the LUNA collaboration of the $d+p\to {}^{3}{\rm He}+\gamma$ rate~\cite{Mossa:2020gjc}, the error budget in the theoretical prediction of the deuterium abundance arises from the lack of detailed knowledge of the rates for $d +d\to n + {}^{3}{\rm He} $ and $d + d \to p + {}^{3}{\rm H}$. For these processes each of the groups uses a slightly different set of rates~\cite{Pisanti:2020efz,Pitrou:2020etk,Yeh:2022heq}, which impacts the theoretical prediction of the deuterium abundance. For a fixed value of $\Omega_b h^2 = 0.02236$, each group reports:
\begin{subequations}\label{eq:DHvals}
\begin{align}
\!\!\!\!\text{D/H}|_{\rm P} &=\left(2.49 \pm 0.08 \right)\times 10^{-5}\,,  \,\, [\text{Yeh et al. 22'}] \\
\!\!\!\!\text{D/H}|_{\rm P} &= \left(2.52\pm 0.07 \right)\times 10^{-5}\,, \,\,  [\text{Pisanti et al. 21'}] \\
\label{eq:DHPitrou}
\!\!\!\!\text{D/H}|_{\rm P} &= \left(2.45 \pm 0.04 \right)\times 10^{-5}\,. \,\,\, [\text{Pitrou et al. 21'}] \end{align}
\end{subequations}
While the results of Yeh et al.~\cite{Yeh:2022heq} and Pisanti et al.~\cite{Pisanti:2020efz} are (within error bars) in good agreement with each other, Pitrou et al.~\cite{Pitrou:2020etk} reports a significantly smaller value. 
In order to assess the impact of this uncertainty in the determination of the primordial lepton asymmetry, we will perform two separate analyses using the rates of Pisanti et al.~\cite{Pisanti:2020efz} (PArthENoPE) and of Pitrou et al.~\cite{Pitrou:2020etk} (PRIMAT).

\begin{figure*}[t]
    \centering
    \begin{tabular}{cc}
    \hspace{0.1cm}    
    \includegraphics[width=0.47\textwidth]{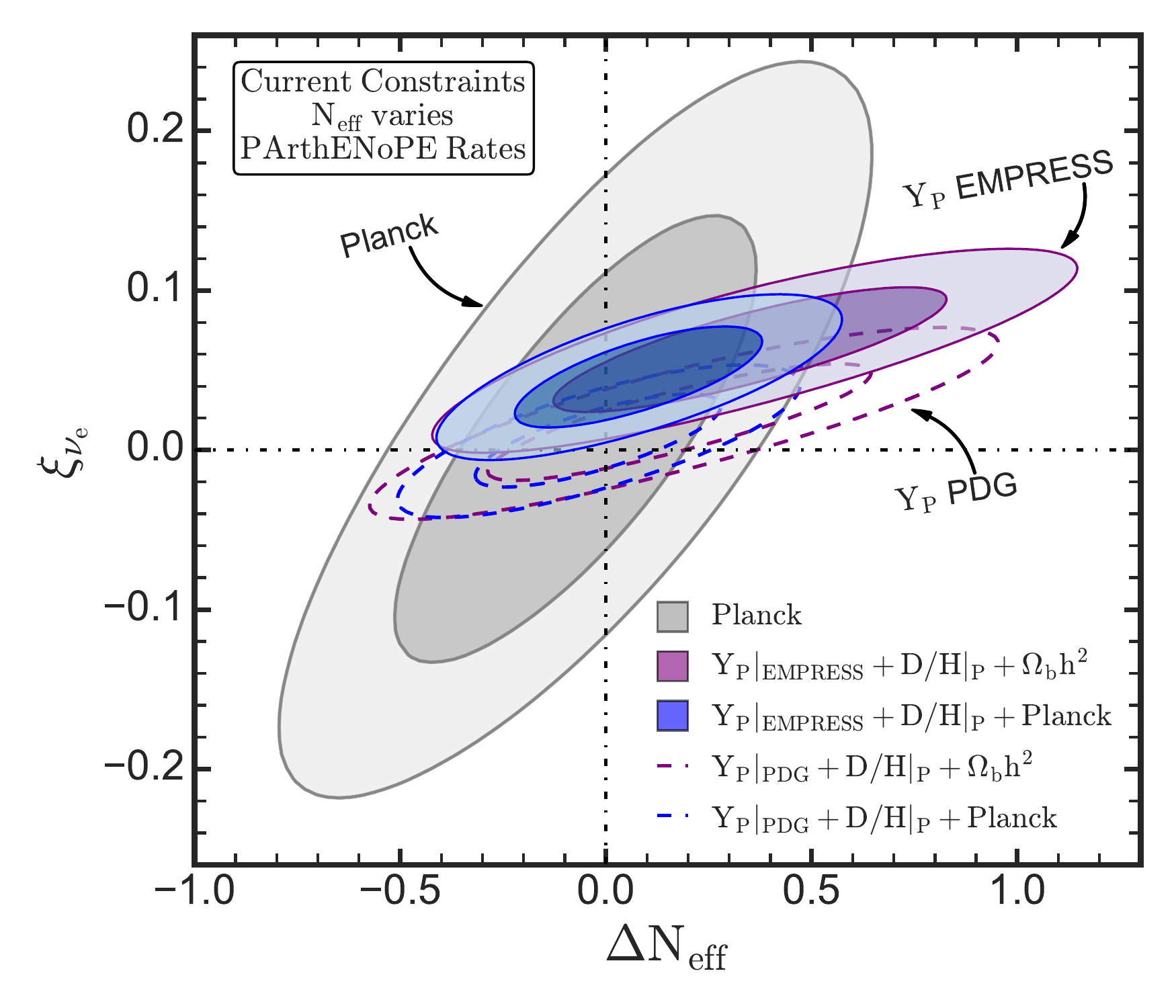}
      &
    \hspace{0.4cm}    
      \includegraphics[width=0.47\textwidth]{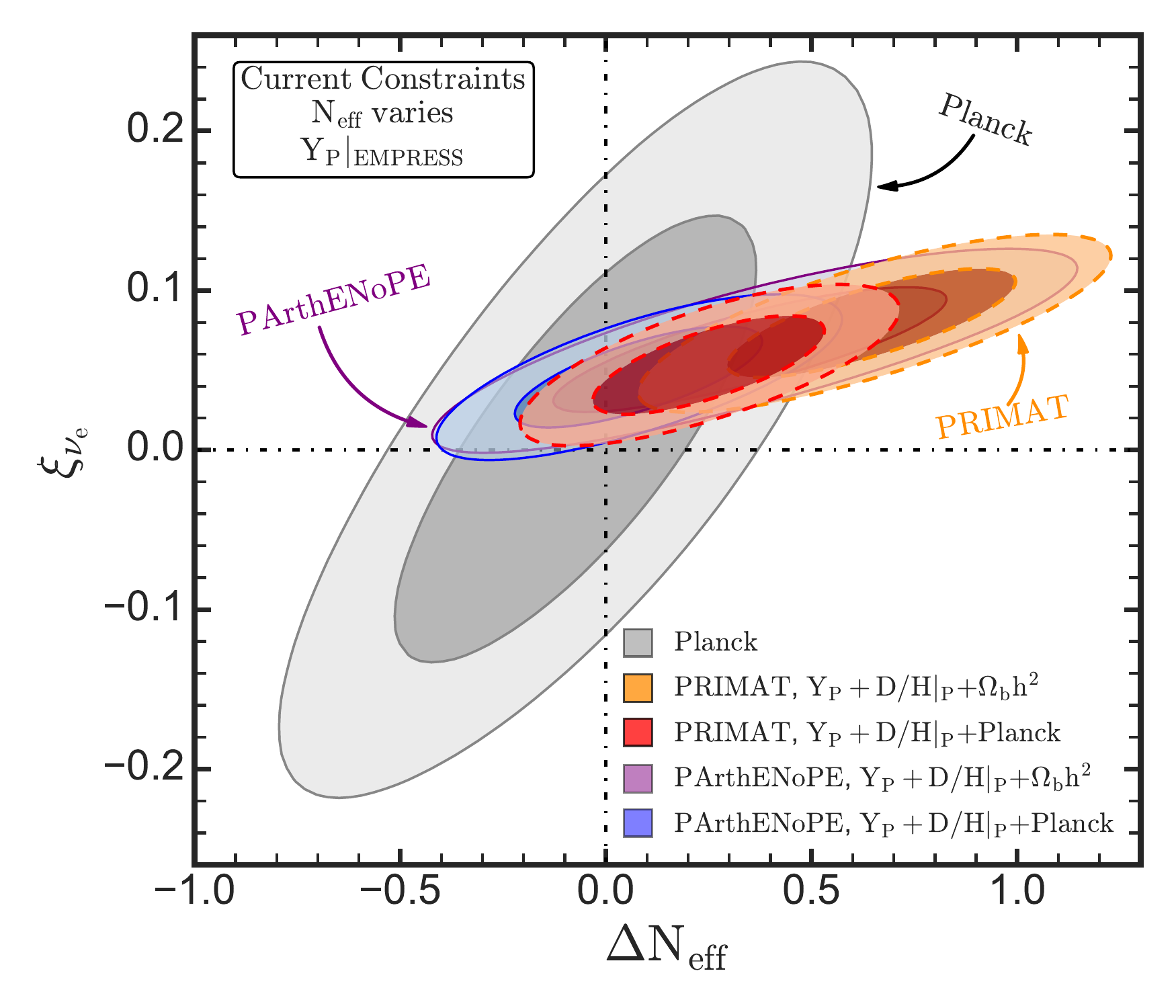}
    \end{tabular}
    \caption{Same as Fig.~\ref{fig:xi_vs_omega_Nefffix}, in the plane of $\xi_{\nu_e}$ and $\Delta N_{\rm eff}$, without making assumptions on the dark radiation content in the Universe.}
    \label{fig:xi_vs_Neff}
\end{figure*}

Our main results are summarized in Figs.~\ref{fig:xi_vs_omega_Nefffix} and \ref{fig:xi_vs_Neff} for cosmological scenarios without and with dark radiation, respectively (see also Table~\ref{tab:CurrentConstraints})\footnote{
We use Gaussian distributions for the different input values and we construct isocontours for the $\Delta\chi^2$ relative to the minimum. In our analysis we also take into account the theoretical uncertainty from the neutron lifetime and the nuclear reaction rates adding in quadrature to the observational uncertainties. Concretely, for helium we take  $\sigma_{\rm Theo}(Y_{\rm P}) = 0.00017$~\cite{Pitrou:2018cgg}, while for the deuterium to hydrogen ratio we take the corresponding values from Eq.~\eqref{eq:DHvals}: $\sigma_{\rm Theo}({\rm D/H}|_{\rm P}) = 0.07\times 10^{-5}$ when using PArthENoPE rates~\cite{Pisanti:2020efz}, and $\sigma_{\rm Theo}({\rm D/H}|_{\rm P}) = 0.04\times 10^{-5}$ when using PRIMAT rates~\cite{Pitrou:2020etk}.}.
In  Fig. ~\ref{fig:xi_vs_omega_Nefffix} we show the 1 and $2\sigma$ confidence regions for $\xi_\nu$ and $\Omega_bh^2$, fixing $N_{\rm eff} = N_{\rm eff}^{\rm SM} =  3.044$.
The left figure shows that current constraints on the (electron) lepton asymmetry $\xi_{\nu_e}$ are dominated by BBN data, and in particular by the primordial helium abundance, with a strong dependence on the value of $Y_{\rm P}$ chosen for the analysis. The new EMPRESS result points to a positive lepton asymmetry,
\begin{align}\label{eq:xinueEMPRESS}
    \xi_{\nu_e} = 0.043\pm 0.015\,\quad [{\rm EMPRESS}]\,,
\end{align}
 which is different from zero with a $\sim 3\sigma$ significance. Instead, if one adopts the PDG-21 recommended value, one obtains:
 \begin{align}
    \xi_{\nu_e} = 0.008\pm 0.013\,\quad [{\rm PDG\!-\!21}]\,,
\end{align}
with no preference for a nonzero lepton asymmetry. The combination with the Planck data does not alter significantly the conclusions for the lepton asymmetry, although it reduces the allowed range for
$\Omega_bh^2$. 

The EMPRESS hint for a nonzero lepton asymmetry is fairly insensitive to the choice of the nuclear reaction rates, as shown in the right panel of Fig. ~\ref{fig:xi_vs_omega_Nefffix}. On the other hand, the reconstructed value of $\Omega_bh^2$ is slightly lower when adopting the PRIMAT rates than for the PArthENoPE rates (see Table~\ref{tab:CurrentConstraints} for a quantitative evaluation of the allowed ranges).

In Fig. ~\ref{fig:xi_vs_Neff} we show the $1$ and $2\sigma$ confidence regions for $\xi_{\nu_e}$ and $\Delta N_{\rm eff}$, corresponding to a scenario with dark radiation. 
The left panel shows that also in this cosmological scenario the determination of $\xi_{\nu_e}$ is dominated by BBN data. On the other hand, the Planck measurements of $N_{\rm eff}$  break the positively correlated degeneracy between $\xi_{\nu_e}$ and $\Delta N_{\rm eff}$, thereby reducing slightly the allowed range of $\xi_{\nu_e}$. As for the scenario without dark radiation, the preferred region of parameter space strongly depends on the value of the primordial helium abundance used in the analysis.  The preferred values of $\xi_{\nu_e}$ and $N_{\rm eff}$, using the EMPRESS determination of $Y_{\rm P}$, are:
 \begin{subequations}\label{eq:EMPRESS_PARTHENOPE}
 \begin{align}
    \xi_{\nu_e} &= 0.046\pm  0.021\,,\quad [{\rm Y_P+D/H|_P+CMB}\,\\
    N_{\rm eff} &= 3.12 \pm 0.20 \,,\qquad \,\,\, {\rm EMPRESS+Planck}]   \label{eq:NeffEMPRESSPart} \,
\end{align}
 \end{subequations}
which amounts to a $2\sigma$ preference for a nonzero lepton asymmetry (see Table~\ref{tab:CurrentConstraints} for a quantitative statement). If one adopts instead the PDG-21 recommended value one finds:
 \begin{subequations}\label{eq:PDG_PARTHENOPE}
 \begin{align}
    \xi_{\nu_e} &= 0.006\pm  0.019\,,\quad [{\rm Y_P+D/H|_P+CMB}\,\\
    N_{\rm eff} &= 3.03 \pm 0.20 \,.\qquad \,\,\, {\rm PDG\!-\!21+Planck}]    \,
\end{align}
 \end{subequations}
yielding no preference for a nonzero lepton asymmetry.

The conclusions on $\xi_{\nu_e}$ do not depend strongly on the choice of the nuclear reaction rates, as shown in the right panel of Fig. ~\ref{fig:xi_vs_Neff}. On the other hand, the preferred values for $\Delta N_{\rm eff}$ can vary sizably depending on this choice. More concretely, using PRIMAT rates and the EMPRESS determination of $Y_{\rm P}$ we find:
 \begin{subequations}\label{eq:EMPRESS_PRIMAT}
 \begin{align}
    \xi_{\nu_e} &= 0.052 \pm 0.020\,,\quad [{\rm Y_P+D/H|_P+CMB}\,\\
    N_{\rm eff} &= 3.29 \pm 0.19 \,,\qquad \,\,\, {\rm EMPRESS+Planck}]    \,
\end{align}
 \end{subequations}
while for the PDG-21 recommended value,
  \begin{subequations}\label{eq:PDG_PRIMAT}
 \begin{align}
    \xi_{\nu_e} &= 0.014 \pm 0.018\,,\quad [{\rm Y_P+D/H|_P+CMB}\,\\
    N_{\rm eff} &= 3.19 \pm 0.18 \,.\qquad \,\,\, {\rm PDG\!-\!21+Planck}]    \,
\end{align}
 \end{subequations}
which should be compared to Eq.~\eqref{eq:EMPRESS_PARTHENOPE} and  Eq.~\eqref{eq:PDG_PARTHENOPE}, respectively. 

It is noteworthy that if one requires $\Delta N_{\rm eff}$ to be positive, as occurs in most models of dark radiation then the preference for a positive lepton asymmetry further increases. We, however, note that in the few cosmological settings that feature $\Delta N_{\rm eff} < 0$, notably MeV-scale reheating~\cite{deSalas:2015glj,Hasegawa:2019jsa} and scenarios with MeV-scale electrophilic particles~\cite{Nollett:2013pwa,Sabti:2019mhn}, these models actually lead to a higher $Y_{\rm P}$, see~\cite{Ichikawa:2005vw,Kolb:1986nf} and would thus enhance the tension with the EMPRESS measurement.

\begin{table*}[t]
\begin{center}
{\def\arraystretch{1.15}
\begin{tabular}{c|c|c|cccc}
\hline\hline
\multicolumn{7}{c}{\textbf{Bounds and Sensitivities on the Primordial Lepton Asymmetries from BBN and CMB data}}     \\ 
\hline \hline
 ${\bf Y_{\rm P}}$ 	&  \textbf{Data Sets}  & \textbf{Nuclear Rates}  & $\qquad \,\,\,\,\,\,\,\, {\bf \xi_{\nu_e}} \,\,\,\,\,\,\,\, \qquad$ & $ \qquad {\bf N_{\rm eff} }\qquad$ & Pref  ${\bf \xi_{\nu_e} \neq 0 }$ & $\,\,\,\chi^2_{\rm min}\,\,\,$ \\
 \hline \hline
 \parbox[t]{8mm}{\multirow{4}{*}{\rotatebox[origin=c]{90}{{CMB}}}}   

& \multirow{4}{*}{{\bf Planck}}                           & PArthENoPE  & $0.022\pm 0.053$ &  3.044  & $0.4\sigma$ & 0  \\ \cline{3-7} 
&   										        & PRIMAT	   & $0.022\pm 0.053$ &  3.044  & $0.4\sigma$ & 0  \\ \cline{3-7} 
&                                                   & PArthENoPE  &$0.004\pm  0.092$ & $ 2.97 \pm 0.29$  & $0.0\sigma$ & 0  \\ \cline{3-7} &   								                & PRIMAT	   &  $0.002\pm0.094$ &  $2.94\pm 0.29$  & $0.0\sigma$ & 0  \\ 
\hline \hline
\parbox[t]{8mm}{\multirow{10}{*}{\rotatebox[origin=c]{90}{{EMPRESS $Y_{\rm P} = 0.2370(34)$}}}}   

& \multirow{2}{*}{{\bf ${\bf Y_{\rm P}}$ + ${{\rm \bf D/H}|_{\rm \bf P}}$}}     & PArthENoPE  & $0.043\pm 0.015$ &  3.044  & $2.9\sigma$ & 0  \\ \cline{3-7} 
&   										            & PRIMAT	   & $0.042\pm 0.015$ &  3.044  & $2.9\sigma$ & 0  \\ \cline{2-7} 
& \multirow{2}{*}{$Y_{\rm P}$ + ${\rm D/H}|_{\rm P}$ + $\Omega_bh^2|_{\rm Planck}$}    & PArthENoPE  & $0.040\pm0.015$ &  3.044  & $2.7\sigma$ & 1.2  \\ \cline{3-7}
&   										                                         & PRIMAT	   & $0.030\pm0.014$ & 3.044  & $2.1\sigma$ & 8.1  \\ \cline{2-7} 
& \multirow{2}{*}{$Y_{\rm P}$ + ${\rm D/H}|_{\rm P}$ + Planck}    & PArthENoPE  & $0.040\pm0.014$ &  3.044  & $2.8\sigma$ & 1  \\ \cline{3-7}
&   										                                         & PRIMAT	   & $0.034\pm0.014$ & 3.044  & $2.4\sigma$ & 7.3  \\ \cline{2-7} 
& \multirow{2}{*}{$Y_{\rm P}$ + ${\rm D/H}|_{\rm P}$ + $\Omega_bh^2|_{\rm Planck}$}    & PArthENoPE  & $0.063\pm  0.026$ & $ 3.39 \pm 0.31$  & $2.4\sigma$ & 0  \\ \cline{3-7} 
&   										                                         & PRIMAT	   & $0.079\pm0.023$ &  $3.68\pm 0.23$  & $3.5\sigma$ & 0  \\ \cline{2-7} 
& \multirow{2}{*}{{\bf ${\bf Y_{\rm P}}$ + ${{\rm \bf D/H}|_{\rm \bf P}}$ + Planck}}  & PArthENoPE  & $0.046\pm  0.021$ & $ 3.12 \pm 0.20$  & $2.2\sigma$ & 0.9  \\ \cline{3-7} 
&   										                  & PRIMAT	   & $0.052\pm0.020$ &  $3.29\pm 0.19$  & $2.6\sigma$ & 5.6  \\ \cline{2-7} 
\hline \hline
\parbox[t]{8mm}{\multirow{10}{*}{\rotatebox[origin=c]{90}{{PDG-21 $Y_{\rm P} = 0.245(3)$}}}}   

& \multirow{2}{*}{{\bf ${\bf Y_{\rm P}}$ + ${{\rm \bf D/H}|_{\rm \bf P}}$ }}     & PArthENoPE  & $0.008\pm 0.013$ &  3.044  & $0.6\sigma$ & 0  \\ \cline{3-7} 
&   										            & PRIMAT	   & $0.007\pm 0.013$ &  3.044  & $0.6\sigma$ & 0  \\ \cline{2-7} 
& \multirow{2}{*}{$Y_{\rm P}$ + ${\rm D/H}|_{\rm P}$ + $\Omega_bh^2|_{\rm Planck}$}    & PArthENoPE  & $0.006\pm0.013$ &  3.044  & $0.5\sigma$ & 0.3  \\ \cline{3-7}
&   										                                         & PRIMAT	   &  $0.000\pm0.013$ & 3.044  & $0.0\sigma$ & 4.4  \\ \cline{2-7} 
& \multirow{2}{*}{$Y_{\rm P}$ + ${\rm D/H}|_{\rm P}$ + Planck}    & PArthENoPE  & $0.008\pm0.013$ &  3.044  & $0.6\sigma$ & 0.4  \\ \cline{3-7}
&   										                                         & PRIMAT	   & $0.004\pm0.013$ & 3.044  & $0.3\sigma$ & 4.9  \\ \cline{2-7} 
& \multirow{2}{*}{$Y_{\rm P}$ + ${\rm D/H}|_{\rm P}$ + $\Omega_bh^2|_{\rm Planck}$}    & PArthENoPE  & $0.018\pm  0.024$ & $ 3.21 \pm 0.31$  & $0.7\sigma$ & 0  \\ \cline{3-7} 
&   										                                         & PRIMAT	   & $0.034\pm0.020$ &  $3.50\pm 0.22$  & $1.7\sigma$ & 0  \\ \cline{2-7} 
& \multirow{2}{*}{{\bf ${\bf Y_{\rm P}}$ + ${{\rm \bf D/H}|_{\rm \bf P}}$ + Planck}}  & PArthENoPE  & $0.006\pm  0.019$ & $ 3.03 \pm 0.20$  & $0.3\sigma$ & 0.5  \\ \cline{3-7} 
&   										                  & PRIMAT	   & $0.014\pm0.018$ &  $3.19\pm 0.18$  & $0.8\sigma$ & 4.3  \\ \cline{2-7} 
\hline \hline
\parbox[t]{8mm}{\multirow{12}{*}{\rotatebox[origin=c]{90}{{Forecasted Constraints}}}}   

& \multirow{2}{*}{$\,\,\,$ {\bf Simons Observatory} ($Y_{\rm P} = 0.2370$) $\,\,\,$}                   & PArthENoPE  & $0.044\pm 0.015$ &  3.044             & $2.9\sigma$ & -  \\ \cline{3-7} 
&   										            & PArthENoPE  & $0.051\pm 0.035$ &  $ 3.13 \pm 0.11$  & $1.4\sigma$ & -  \\ \cline{2-7} 
& \multirow{2}{*}{{\bf CMB-S4} ($Y_{\rm P} = 0.2370$)}                               & PArthENoPE  & $0.044\pm 0.010$ &  3.044             & $4.2\sigma$ & -  \\ \cline{3-7} 
&   										            & PArthENoPE  & $0.051\pm 0.023$ &  $ 3.13 \pm 0.08$  & $2.1\sigma$ & -  \\ \cline{2-7}
& \multirow{2}{*}{Simons Observatory + EMPRESS}          & PArthENoPE  & $0.043\pm 0.010$ &  3.044             & $4.4\sigma$ & -  \\ \cline{3-7} 
&   										            & PArthENoPE  & $0.047\pm 0.016$ &  $ 3.12 \pm 0.07$  & $2.9\sigma$ & -  \\ \cline{2-7} 
& \multirow{2}{*}{CMB-S4 + EMPRESS}                     & PArthENoPE  & $0.043\pm 0.008$ &  3.044             & $5.3\sigma$ & -  \\ \cline{3-7} 
&   										            & PArthENoPE  & $0.045\pm 0.014$ &  $ 3.12 \pm 0.06$  & $3.3\sigma$ & -  \\ \cline{2-7}
& \multirow{2}{*}{Simons Observatory + $Y_{\rm P}$ SM}    & PArthENoPE  & $\!\!\!\!-0.001\pm 0.010$ &  3.044             & $0.0\sigma$ & -  \\ \cline{3-7} 
&   										            & PArthENoPE  & $0.001\pm 0.015$ &  $ 3.05 \pm 0.07$  & $0.1\sigma$ & -  \\ \cline{2-7} 
& \multirow{2}{*}{CMB-S4 + $Y_{\rm P}$ SM}                & PArthENoPE  & $0.000\pm 0.008$ &  3.044             & $0.0\sigma$ & -  \\ \cline{3-7} 
&   										            & PArthENoPE  & $0.001\pm 0.013$ &  $ 3.05 \pm 0.06$  & $0.0\sigma$ & -  \\ \cline{2-7}
\hline \hline
\end{tabular}
}
\end{center}
\vspace{-0.3cm}
\caption{Summary of constraints or forecasts on the primordial (electron) lepton asymmetry, $\xi_{\nu_e}$, from considering several combinations of BBN and CMB data, for cosmological scenarios without or with dark radiation, and for two possible choices of the nuclear reaction rates. See main text for details.}
\label{tab:CurrentConstraints}
\end{table*}

\section{Forecasts for the Simons Observatory and CMB-S4}\label{sec:forecast}

Future CMB observations will be instrumental to further probe the hint for a nonzero lepton asymmetry from EMPRESS. The reason is twofold. First, they will provide an independent and precise measurement of $Y_{\rm P}$, and second, they will yield an unprecedented sensitivity to $N_{\rm eff}$ which, as shown  e.g. in Fig. ~\ref{fig:xi_vs_Neff}, is positively correlated with $\xi_{\nu_e}$. In this section we consider specifically the prospects for detecting a nonzero primordial asymmetry with the upcoming Simons Observatory and the projected CMB-S4. 

To this end, we take the baseline covariance matrix from the Simons Observatory to the relevant parameters of our analysis $Y_{\rm P}$, $N_{\rm eff}$ and $\Omega_bh^2$~\cite{SimonsObservatory:2018koc}. Once marginalized over the rest of cosmological parameters, they read~\cite{Sabti:2019mhn}:
\begin{subequations}\label{eq:YPSimons_Neff}
\begin{align}
{\rm Simons}\,&\, {\rm Observatory}\nonumber \\
\sigma(\Omega_b h^2) &= 0.000073 \,,\\
\sigma(Y_{\rm P}) &= 0.0066 \,,\\
\sigma(N_{\rm eff}) &= 0.11 \,,\\
\rho(\Omega_b h^2,\,Y_{\rm P})   &= 0.33\,,\\
\rho(\Omega_b h^2,\,N_{\rm eff}) &= 0.072\,,\\
\rho(N_{\rm eff},\,Y_{\rm P})  &= -0.86\,.
\end{align}
\end{subequations}
For CMB-S4, we use the results from the Fisher matrix forecast performed in~\cite{Sabti:2019mhn} which is in very good agreement with the results reported  by the collaboration~\cite{CMB-S4:2016ple,Abazajian:2019eic}. The relevant parameters read:
\begin{subequations}\label{eq:YPCMB-S4_Neff}
\begin{align}
{\rm CMB}&{\rm -S4}\nonumber \\
\sigma(\Omega_b h^2) &= 0.000047 \,,\\
\sigma(Y_{\rm P}) &= 0.0043 \,,\\
\sigma(N_{\rm eff}) &= 0.081 \,,\\
\rho(\Omega_b h^2,\,Y_{\rm P})   &= 0.22\,,\\
\rho(\Omega_b h^2,\,N_{\rm eff}) &= 0.25\,,\\
\rho(N_{\rm eff},\,Y_{\rm P})  &= -0.84\,.
\end{align}
\end{subequations}
For the central value of the baryon density we will take  $\Omega_bh^2 = 0.02242$, as favored by Planck CMB observations, see Eq.~\eqref{eq:omegabprior}. 
For $Y_{\rm P}$ we will consider two possibilities, either   $Y_{\rm P} = Y_{\rm P}|_{\rm SBBN}= 0.2469$ or $Y_{\rm P} = Y_{\rm P}|_{\rm EMPRESS}=0.2370$, in order to make forecasts for the cases where the helium abundance coincides with the standard BBN prediction, or when it is lower as hinted by EMPRESS. For both, we consider also a direct astrophysical determination with an error bar of $0.003$ which matches the precision of current determinations.
Finally, for $N_{\rm eff}$ we will either choose $N_{\rm eff}^{\rm SM} = 3.044$, as expected in the Standard Model, or the central value inferred from the current full analysis of BBN and CMB data using PArthENoPE rates, namely $N_{\rm eff} = 3.12$, see Eq.~\eqref{eq:NeffEMPRESSPart}.

\begin{figure*}[t]
    \centering
    \begin{tabular}{cc}
      \includegraphics[width=0.5\textwidth]{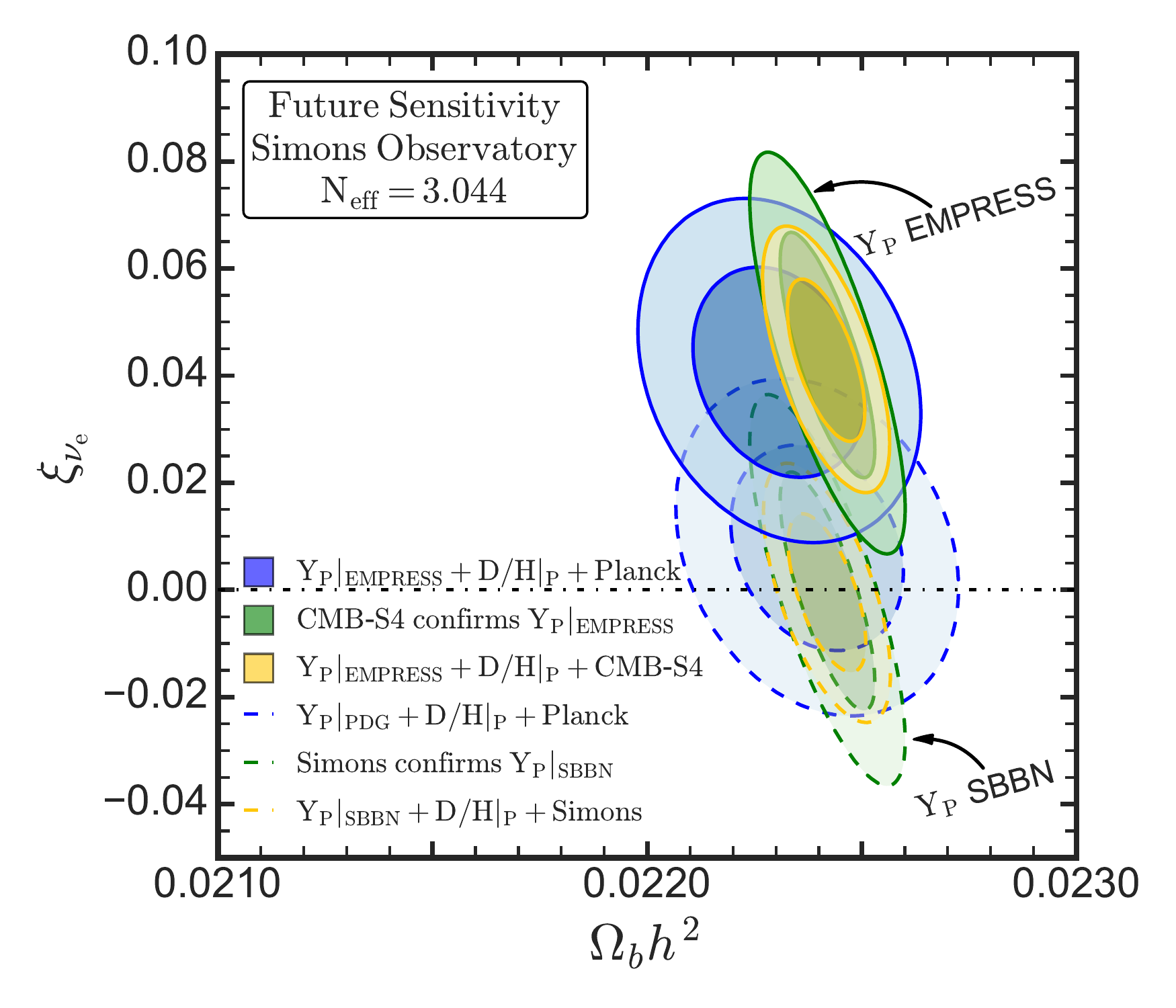}
      &
      \includegraphics[width=0.5\textwidth]{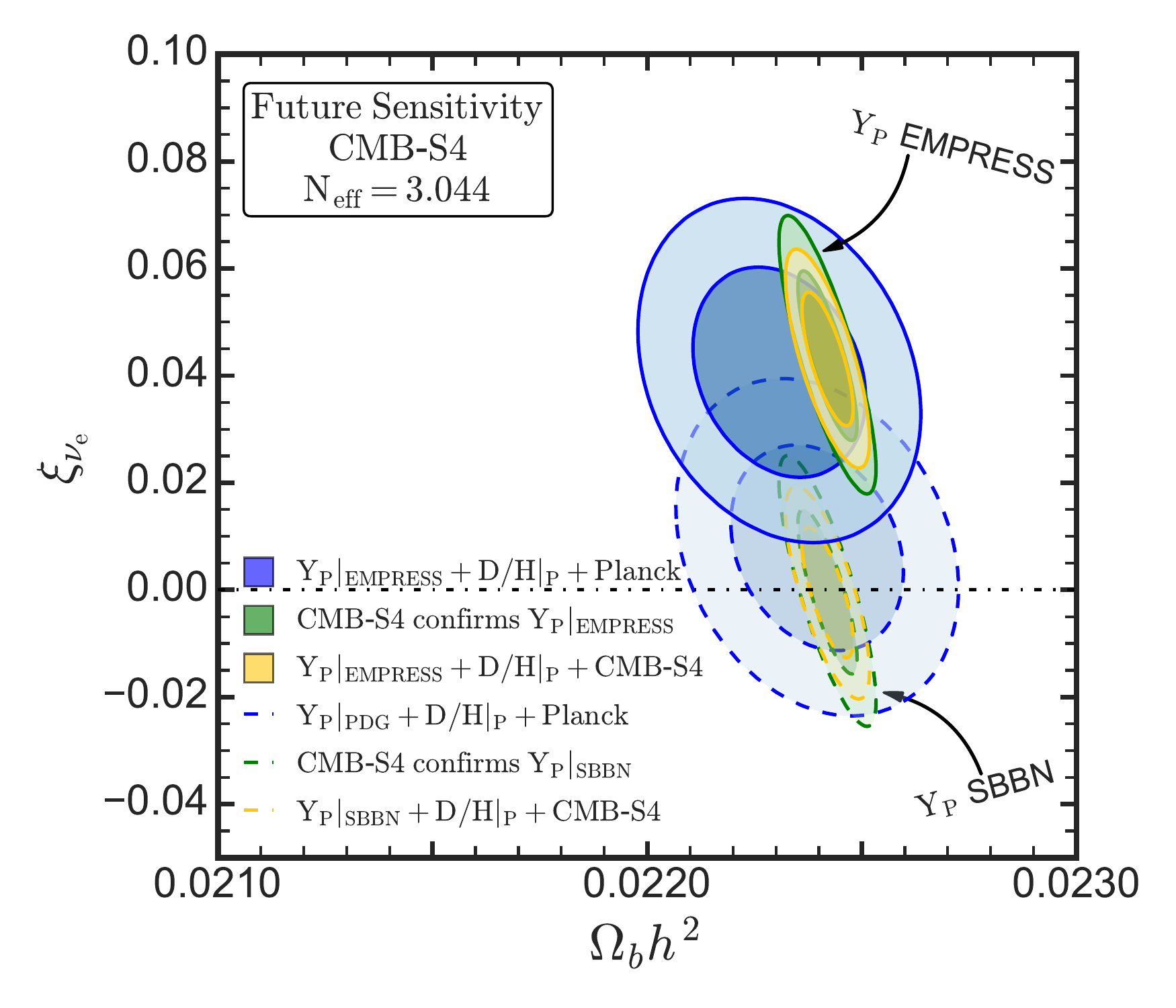}\\
 \hspace{-0.5cm}   \includegraphics[width=0.48\textwidth]{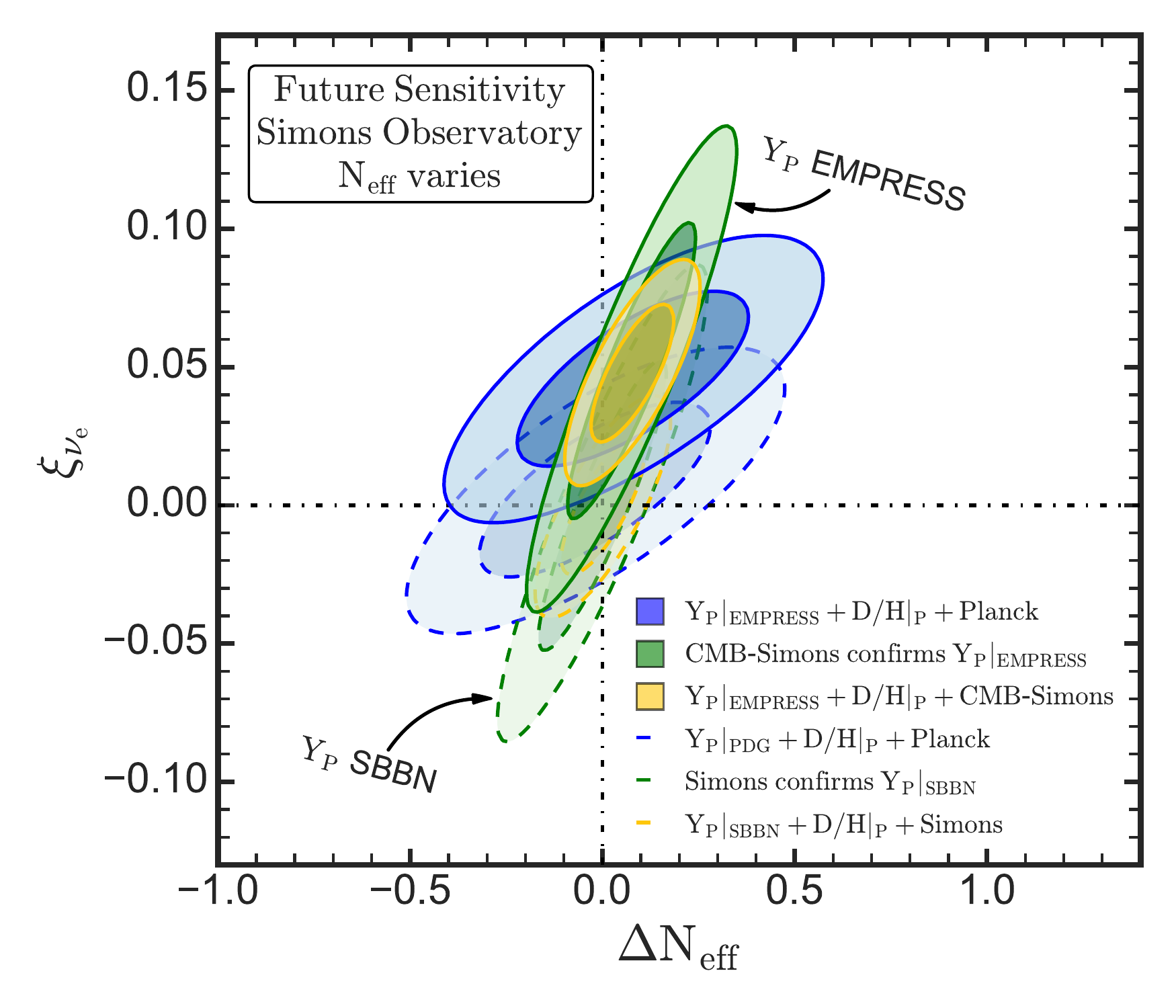}
      &
  \hspace{-0.5cm}  \includegraphics[width=0.48\textwidth]{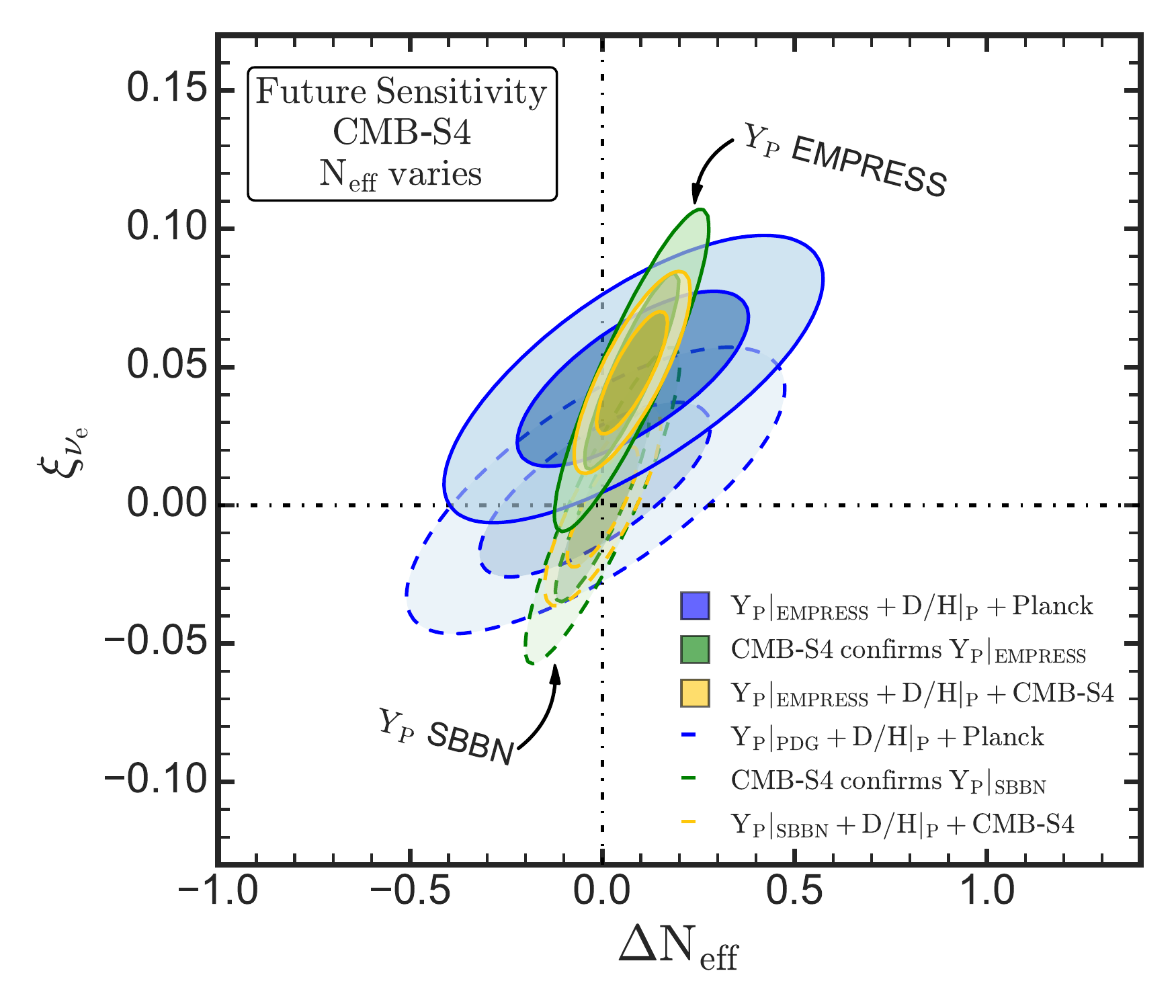}
    \end{tabular}
    \caption{ $1$ and $2\sigma$ C.L. forecast regions for $\xi_{\nu_e}$ and $\Omega_bh^2$ for a scenario without dark radiation (top panels), or $\xi_{\nu_e}$ and $\Delta N_{\rm eff}$ for a scenario without making assumptions on the amount of dark radiation (bottom panels) from nucleosynthesis data, the upcoming Simons Observatory (left panels) or the projected CMB-S4 (right panels), and their combination.
    }
    \label{fig:forecast}
\end{figure*}

In Fig. ~\ref{fig:forecast} we present the results of our forecast, taking for concreteness the  PArthENoPE rates (the results for the PRIMAT rates are practically identical).
In the upper panels of Fig. ~\ref{fig:forecast} we show the sensitivity to $\xi_{\nu_e}$ from the Simons Observatory (left) or CMB-S4 (right) as a function of $\Omega_bh^2$ for a scenario with a fixed $N_{\rm eff}=3.044$. We compare this sensitivity to the one obtained from current CMB+BBN data. We note that the Simons Observatory on its own has the power to reach a sensitivity to $\xi_{\nu_e}$ that will be competitive with current combined constraints. 
Furthermore, we find that CMB-S4 will improve significantly upon the Simons Observatory.
More concretely, our forecast sensitivity to the lepton asymmetry for each experiments reads:
\begin{subequations}
\begin{align}
    \sigma(\xi_{\nu_e})|_{N_{\rm eff} = 3.044} &\simeq  0.015\,,  \quad [{\rm Simons \,\, Obs.}]\\ 
    \sigma(\xi_{\nu_e})|_{N_{\rm eff} = 3.044} &\simeq  0.010\,. \,\quad [{\rm CMB\!-\!S4}]
\end{align}
\end{subequations}
More importantly, if the true value of the helium abundance correspond to the EMPRESS central value, $Y_{\rm P} =0.2370$, and the Universe does not contain substantial amounts of dark radiation, $N_{\rm eff}=3.044$, then the combination of EMPRESS and the Simons Observatory would increase the significance for a nonzero lepton asymmetry, to $\sim 4.4\sigma$, and the combination with CMB-S4 to  $\sim 5.3\sigma$ (see  Table~\ref{tab:CurrentConstraints}).

In the lower panels of Fig. ~\ref{fig:forecast}, we leave $N_{\rm eff}$ as an unconstrained parameter. As expected, the reach of the Simons Observatory and of CMB-S4 worsen when relaxing the assumptions on the cosmological scenario. We obtain:
\begin{subequations}
\begin{align}
    \sigma(\xi_{\nu_e})   &\simeq  0.04\,, \quad[{\rm Simons \,\, Obs.}]\\ 
    \sigma(N_{\rm eff}) &\simeq  0.11\,, \quad [{\rm Simons \,\, Obs.}]\\ 
    \sigma(\xi_{\nu_e})   &\simeq  0.02 \,, \quad [{\rm CMB\!-\!S4}]\\ 
    \sigma(N_{\rm eff}) &\simeq  0.08\,.\, \quad [{\rm CMB\!-\!S4}]
\end{align}
\end{subequations}

\begin{figure*}[t]
    \centering
    \begin{tabular}{cc}
\hspace{-0.6cm}      \includegraphics[width=0.5\textwidth]{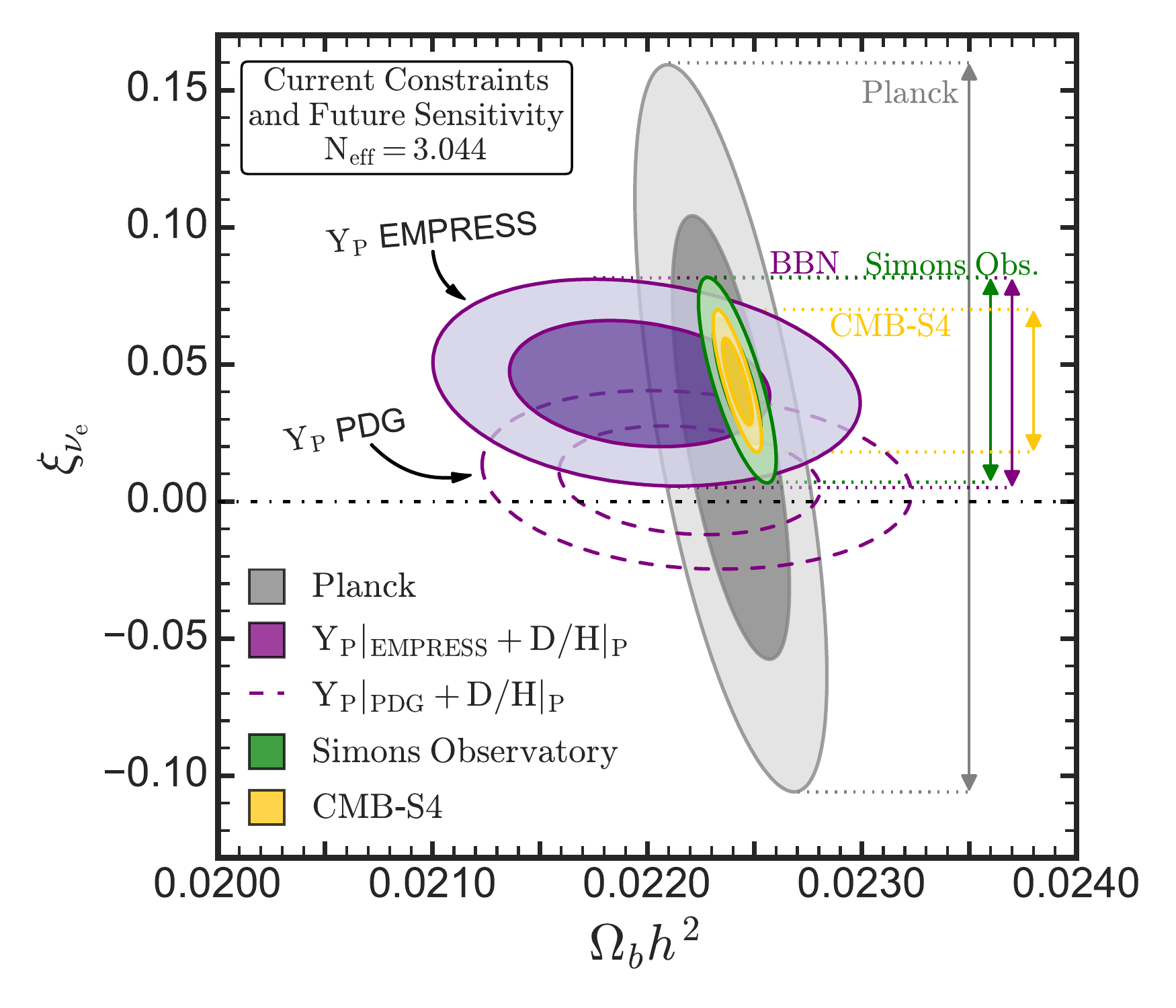}
      &
      \includegraphics[width=0.5\textwidth]{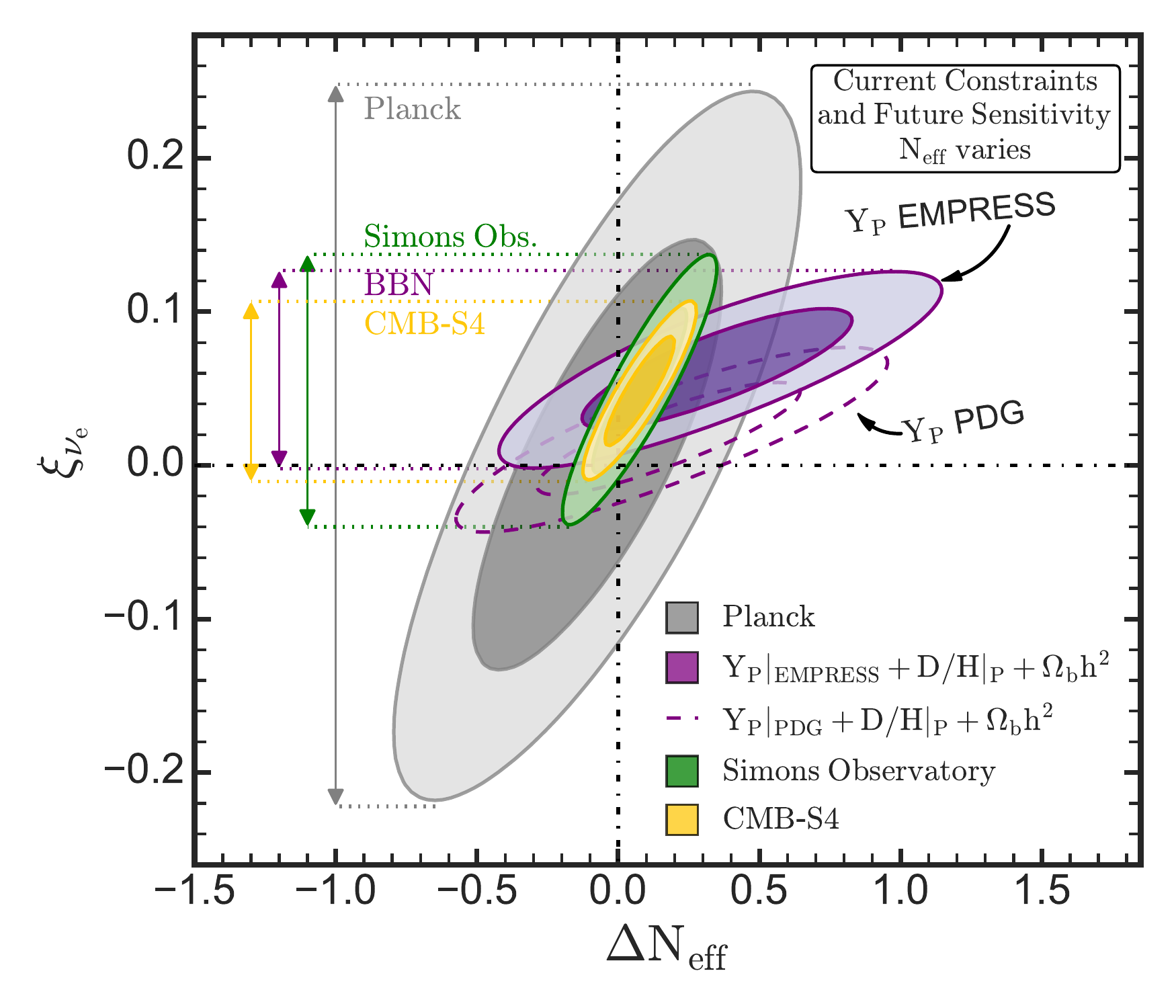}
    \end{tabular}
    \vspace{-0.4cm}
    \caption{Summary of current and forecast 1- and 2-$\sigma$ allowed regions for $\xi_{\nu_e}$ and $\Omega_bh^2$ in a scenario without dark radiation (left panel), or $\xi_{\nu_e}$ and $\Delta N_{\rm eff}$ in a scenario without making assumptions on the amount of dark radiation (right panel) from nucleosynthesis data (EMPRESS survey or PDG-21 recommended value) and CMB data (Planck, Simons Observatory or CMB-S4). }
    \label{fig:plotintro}
\end{figure*}

Yet, the combination of EMPRESS with CMB experiments will significantly narrow down the allowed ranges for $\xi_{\nu_e}$ and $\Delta N_{\rm eff}$, and would strengthen the case for a nonzero lepton asymmetry, should the EMPRESS hint be correct. Concretely, while current data only give a $2\sigma$ significance for a nonzero lepton asymmetry (when leaving $N_{\rm eff}$ unconstrained), the combination with the Simons Observatory or CMB-S4 would increase the significance to $\sim 3\sigma$. Concretely, we obtain
 \begin{subequations}\label{eq:EMPRESS_CMB_forecast}
 \begin{align}
  \!\!\!\!\! \xi_{\nu_e} &= 0.047 \pm 0.016\,,\,\,\,\, [{\rm EMPRESS+Simons Obs.}]    \\
    \!\!\!\!\!   N_{\rm eff} &= 3.12 \pm 0.07 \,,\,\,\,\,\,\,\,\,\,\, [{\rm EMPRESS+Simons Obs.}]   \\
    \!\!\!\!\!    \xi_{\nu_e} &= 0.045 \pm 0.014\,,\,\,\,\, [{\rm EMPRESS+CMB\!-\!S4}]    \\
      \!\!\!\!\!  N_{\rm eff} &= 3.12 \pm 0.06 \,.\,\,\,\,\,\,\,\,\,\,\, [{\rm EMPRESS+CMB\!-\!S4}]   
\end{align}
 \end{subequations}

Let us finalize this section commenting on the possible role of the primordial deuterium abundance as a third (independent) probe of a primordial lepton asymmetry, along with the CMB and the helium data. The current measurement is limited by statistics, however it is expected to improve substantially in the near future with the advent of 30m class optical/near-infrared telescopes~\cite{Grohs:2019cae}. On the other hand, 
the theoretical prediction for ${\rm D/H}|_{\rm P}$ is currently limited by uncertainties in the $d +d\to n + {}^{3}{\rm He} $ and $d + d \to p + {}^{3}{\rm H}$ reaction rates. Therefore, in order to provide a competitive probe of the lepton asymmetry, it is mandatory to measure more precisely these reactions, or improve the theoretical modeling~\cite{Pitrou:2021vqr}.

\section{Conclusions}\label{sec:conclusions}

The recent measurement of the primordial helium abundance by EMPRESS could be an indication for a nonzero lepton asymmetry in the electron neutrino flavor. Motivated by this new measurement, we have performed a global analysis of the primordial lepton asymmetries using both BBN and CMB data. Our main results are summarized in Fig. ~\ref{fig:plotintro}, which shows the current constraints on the lepton asymmetry (parametrized by the neutrino chemical potential $\xi_{\nu_e}$) and its correlation with the baryon asymmetry ($\Omega_bh^2$) and with the amount of dark radiation in the Universe (parametrized by the extra contributions to the effective number of neutrino species,  $\Delta N_{\rm eff}$); quantitative results are reported in  Table~\ref{tab:CurrentConstraints}.

We have found that the determination of the lepton asymmetry is currently dominated by the helium abundance, and is strongly dependent on the dataset considered, ranging from a $\sim 3\sigma$ indication for a nonzero lepton asymmetry when using the EMPRESS data, to no significant indication when using the PDG-21 recommended value (see Fig. ~\ref{fig:xi_vs_omega_Nefffix} and Table~\ref{tab:CurrentConstraints}).
Our conclusions are in agreement with other recent works also analyzing the implications of the EMPRESS measurements on the cosmological parameters~\cite{Matsumoto:2022tlr,Burns:2022hkq}.

Further, we have also investigated the impact of the uncertainties in the nuclear reaction rates for the determination of the lepton asymmetry, taking specifically the rates from  PArthENoPE and from PRIMAT. We have concluded that the choice of nuclear reaction rates does not affect significantly the determination of the lepton asymmetry, both when $N_{\rm eff}$ is fixed and when it is allowed to float.

Finally, we have also performed a forecast of the sensitivity to the lepton asymmetry from the upcoming Simons Observatory and the future CMB-S4. These experiments, by themselves, will have a sensitivity to the lepton asymmetry which is comparable to our current global fit. Should the helium abundance be lower than the SM prediction, the CMB data from the Simons Observatory, combined with the results from EMPRESS, will strengthen the hint for a nonzero lepton asymmetry to $\sim 3\sigma$ if no assumption is done on the cosmological parameters, and $\sim 4\sigma$ if it is assumed that the Universe does not contain dark radiation. With the future CMB-S4 data the significance would increase to $\sim 5\sigma$. 

If confirmed, this result would hint toward new physics generating a lepton asymmetry at low temperatures, to prevent its conversion into a baryon asymmetry by sphaleron processes. The construction of possible models and their possible signals deserves in our opinion further investigation. 

\begin{center}\textbf{Acknowledgments}  \end{center}

\noindent This work was supported by the Collaborative Research Center SFB1258 and by the Deutsche Forschungsgemeinschaft (DFG, German Research Foundation) under Germany's Excellence
Strategy - EXC-2094 - 390783311.  M.E. is supported by a Fellowship of the Alexander von Humboldt Foundation. We gratefully acknowledge the use of the public BBN code {\tt PArthENoPE-v3.0}~\cite{Pisanti:2007hk,Consiglio:2017pot,Gariazzo:2021iiu}.

\bibliography{biblio.bib}

\providecommand{\href}[2]{#2}\begingroup\raggedright\begin{thebibliography}{10}

\bibitem{Kolb:1990vq}
E.~W. Kolb and M.~S. Turner, \emph{{The Early Universe}}, {\emph{Front. Phys.}
  {\bfseries 69} (1990) 1}.

\bibitem{Gorbunov:2011zz}
V.~A. Rubakov and D.~S. Gorbunov, \emph{{Introduction to the Theory of the
  Early Universe}: {Hot big bang theory}}. World Scientific, Singapore, 2017,
  \href{https://doi.org/10.1142/10447}{10.1142/10447}.

\bibitem{planck}
{\scshape Planck} collaboration, N.~Aghanim et~al., \emph{{Planck 2018 results.
  VI. Cosmological parameters}},
  \href{https://doi.org/10.1051/0004-6361/201833910}{\emph{Astron. Astrophys.}
  {\bfseries 641} (2020) A6}
  [\href{https://arxiv.org/abs/1807.06209}{{\ttfamily 1807.06209}}].

\bibitem{Mossa:2020gjc}
V.~Mossa et~al., \emph{{The baryon density of the Universe from an improved
  rate of deuterium burning}},
  \href{https://doi.org/10.1038/s41586-020-2878-4}{\emph{Nature} {\bfseries
  587} (2020) 210}.

\bibitem{Pisanti:2020efz}
O.~Pisanti, G.~Mangano, G.~Miele and P.~Mazzella, \emph{{Primordial Deuterium
  after LUNA: concordances and error budget}},
  \href{https://doi.org/10.1088/1475-7516/2021/04/020}{\emph{JCAP} {\bfseries
  04} (2021) 020} [\href{https://arxiv.org/abs/2011.11537}{{\ttfamily
  2011.11537}}].

\bibitem{Pitrou:2020etk}
C.~Pitrou, A.~Coc, J.-P. Uzan and E.~Vangioni, \emph{{A new tension in the
  cosmological model from primordial deuterium?}},
  \href{https://doi.org/10.1093/mnras/stab135}{\emph{Mon. Not. Roy. Astron.
  Soc.} {\bfseries 502} (2021) 2474}
  [\href{https://arxiv.org/abs/2011.11320}{{\ttfamily 2011.11320}}].

\bibitem{Yeh:2022heq}
T.-H. Yeh, J.~Shelton, K.~A. Olive and B.~D. Fields, \emph{{Probing physics
  beyond the standard model: limits from BBN and the CMB independently and
  combined}}, \href{https://doi.org/10.1088/1475-7516/2022/10/046}{\emph{JCAP}
  {\bfseries 10} (2022) 046}
  [\href{https://arxiv.org/abs/2207.13133}{{\ttfamily 2207.13133}}].

\bibitem{Kuzmin:1985mm}
V.~A. Kuzmin, V.~A. Rubakov and M.~E. Shaposhnikov, \emph{{On the Anomalous
  Electroweak Baryon Number Nonconservation in the Early Universe}},
  \href{https://doi.org/10.1016/0370-2693(85)91028-7}{\emph{Phys. Lett. B}
  {\bfseries 155} (1985) 36}.

\bibitem{Khlebnikov:1988sr}
S.~Y. Khlebnikov and M.~E. Shaposhnikov, \emph{{The Statistical Theory of
  Anomalous Fermion Number Nonconservation}},
  \href{https://doi.org/10.1016/0550-3213(88)90133-2}{\emph{Nucl. Phys. B}
  {\bfseries 308} (1988) 885}.

\bibitem{Harvey:1990qw}
J.~A. Harvey and M.~S. Turner, \emph{{Cosmological baryon and lepton number in
  the presence of electroweak fermion number violation}},
  \href{https://doi.org/10.1103/PhysRevD.42.3344}{\emph{Phys. Rev. D}
  {\bfseries 42} (1990) 3344}.

\bibitem{Dreiner:1992vm}
H.~K. Dreiner and G.~G. Ross, \emph{{Sphaleron erasure of primordial
  baryogenesis}},
  \href{https://doi.org/10.1016/0550-3213(93)90579-E}{\emph{Nucl. Phys. B}
  {\bfseries 410} (1993) 188}
  [\href{https://arxiv.org/abs/hep-ph/9207221}{{\ttfamily hep-ph/9207221}}].

\bibitem{Casas:1997gx}
A.~Casas, W.~Y. Cheng and G.~Gelmini, \emph{{Generation of large lepton
  asymmetries}},
  \href{https://doi.org/10.1016/S0550-3213(98)00606-3}{\emph{Nucl. Phys. B}
  {\bfseries 538} (1999) 297}
  [\href{https://arxiv.org/abs/hep-ph/9709289}{{\ttfamily hep-ph/9709289}}].

\bibitem{Dolgov:1989us}
A.~D. Dolgov and D.~P. Kirilova, \emph{{ON PARTICLE CREATION BY A TIME
  DEPENDENT SCALAR FIELD}}, {\emph{Sov. J. Nucl. Phys.} {\bfseries 51} (1990)
  172}.

\bibitem{Bajc:1997ky}
B.~Bajc, A.~Riotto and G.~Senjanovic, \emph{{Large lepton number of the
  universe and the fate of topological defects}},
  \href{https://doi.org/10.1103/PhysRevLett.81.1355}{\emph{Phys. Rev. Lett.}
  {\bfseries 81} (1998) 1355}
  [\href{https://arxiv.org/abs/hep-ph/9710415}{{\ttfamily hep-ph/9710415}}].

\bibitem{Asaka:2005pn}
T.~Asaka and M.~Shaposhnikov, \emph{{The $\nu$MSM, dark matter and baryon
  asymmetry of the universe}},
  \href{https://doi.org/10.1016/j.physletb.2005.06.020}{\emph{Phys. Lett. B}
  {\bfseries 620} (2005) 17}
  [\href{https://arxiv.org/abs/hep-ph/0505013}{{\ttfamily hep-ph/0505013}}].

\bibitem{Asaka:2005an}
T.~Asaka, S.~Blanchet and M.~Shaposhnikov, \emph{{The nuMSM, dark matter and
  neutrino masses}},
  \href{https://doi.org/10.1016/j.physletb.2005.09.070}{\emph{Phys. Lett. B}
  {\bfseries 631} (2005) 151}
  [\href{https://arxiv.org/abs/hep-ph/0503065}{{\ttfamily hep-ph/0503065}}].

\bibitem{Pilaftsis:2003gt}
A.~Pilaftsis and T.~E.~J. Underwood, \emph{{Resonant leptogenesis}},
  \href{https://doi.org/10.1016/j.nuclphysb.2004.05.029}{\emph{Nucl. Phys. B}
  {\bfseries 692} (2004) 303}
  [\href{https://arxiv.org/abs/hep-ph/0309342}{{\ttfamily hep-ph/0309342}}].

\bibitem{Borah:2022uos}
D.~Borah and A.~Dasgupta, \emph{{Large Neutrino Asymmetry from TeV Scale
  Leptogenesis in the Light of Helium Anomaly}},
  \href{https://arxiv.org/abs/2206.14722}{{\ttfamily 2206.14722}}.

\bibitem{Kawasaki:2002hq}
M.~Kawasaki, F.~Takahashi and M.~Yamaguchi, \emph{{Large lepton asymmetry from
  Q balls}}, \href{https://doi.org/10.1103/PhysRevD.66.043516}{\emph{Phys. Rev.
  D} {\bfseries 66} (2002) 043516}
  [\href{https://arxiv.org/abs/hep-ph/0205101}{{\ttfamily hep-ph/0205101}}].

\bibitem{Kawasaki:2022hvx}
M.~Kawasaki and K.~Murai, \emph{{Lepton asymmetric universe}},
  \href{https://doi.org/10.1088/1475-7516/2022/08/041}{\emph{JCAP} {\bfseries
  08} (2022) 041} [\href{https://arxiv.org/abs/2203.09713}{{\ttfamily
  2203.09713}}].

\bibitem{March-Russell:1999hpw}
J.~March-Russell, H.~Murayama and A.~Riotto, \emph{{The Small observed baryon
  asymmetry from a large lepton asymmetry}},
  \href{https://doi.org/10.1088/1126-6708/1999/11/015}{\emph{JHEP} {\bfseries
  11} (1999) 015} [\href{https://arxiv.org/abs/hep-ph/9908396}{{\ttfamily
  hep-ph/9908396}}].

\bibitem{Domcke:2022uue}
V.~Domcke, K.~Kamada, K.~Mukaida, K.~Schmitz and M.~Yamada, \emph{{A new
  constraint on primordial lepton flavour asymmetries}},
  \href{https://arxiv.org/abs/2208.03237}{{\ttfamily 2208.03237}}.

\bibitem{Sarkar:1995dd}
S.~Sarkar, \emph{{Big bang nucleosynthesis and physics beyond the standard
  model}}, \href{https://doi.org/10.1088/0034-4885/59/12/001}{\emph{Rept. Prog.
  Phys.} {\bfseries 59} (1996) 1493}
  [\href{https://arxiv.org/abs/hep-ph/9602260}{{\ttfamily hep-ph/9602260}}].

\bibitem{Iocco:2008va}
F.~Iocco, G.~Mangano, G.~Miele, O.~Pisanti and P.~D. Serpico, \emph{{Primordial
  Nucleosynthesis: from precision cosmology to fundamental physics}},
  \href{https://doi.org/10.1016/j.physrep.2009.02.002}{\emph{Phys. Rept.}
  {\bfseries 472} (2009) 1} [\href{https://arxiv.org/abs/0809.0631}{{\ttfamily
  0809.0631}}].

\bibitem{Pitrou:2018cgg}
C.~Pitrou, A.~Coc, J.-P. Uzan and E.~Vangioni, \emph{{Precision big bang
  nucleosynthesis with improved Helium-4 predictions}},
  \href{https://doi.org/10.1016/j.physrep.2018.04.005}{\emph{Phys. Rept.}
  {\bfseries 754} (2018) 1} [\href{https://arxiv.org/abs/1801.08023}{{\ttfamily
  1801.08023}}].

\bibitem{Lesgourgues:2013sjj}
J.~Lesgourgues, G.~Mangano, G.~Miele and S.~Pastor, \emph{{Neutrino
  Cosmology}}. Cambridge University Press, 2, 2013.

\bibitem{Serpico:2005bc}
P.~D. Serpico and G.~G. Raffelt, \emph{{Lepton asymmetry and primordial
  nucleosynthesis in the era of precision cosmology}},
  \href{https://doi.org/10.1103/PhysRevD.71.127301}{\emph{Phys. Rev. D}
  {\bfseries 71} (2005) 127301}
  [\href{https://arxiv.org/abs/astro-ph/0506162}{{\ttfamily
  astro-ph/0506162}}].

\bibitem{Mangano:2011ip}
G.~Mangano, G.~Miele, S.~Pastor, O.~Pisanti and S.~Sarikas, \emph{{Updated BBN
  bounds on the cosmological lepton asymmetry for non-zero $\theta_{13}$}},
  \href{https://doi.org/10.1016/j.physletb.2012.01.015}{\emph{Phys. Lett. B}
  {\bfseries 708} (2012) 1} [\href{https://arxiv.org/abs/1110.4335}{{\ttfamily
  1110.4335}}].

\bibitem{Chu:2006ua}
Y.-Z. Chu and M.~Cirelli, \emph{{Sterile neutrinos, lepton asymmetries,
  primordial elements: How much of each?}},
  \href{https://doi.org/10.1103/PhysRevD.74.085015}{\emph{Phys. Rev. D}
  {\bfseries 74} (2006) 085015}
  [\href{https://arxiv.org/abs/astro-ph/0608206}{{\ttfamily
  astro-ph/0608206}}].

\bibitem{Simha:2008mt}
V.~Simha and G.~Steigman, \emph{{Constraining The Universal Lepton Asymmetry}},
  \href{https://doi.org/10.1088/1475-7516/2008/08/011}{\emph{JCAP} {\bfseries
  08} (2008) 011} [\href{https://arxiv.org/abs/0806.0179}{{\ttfamily
  0806.0179}}].

\bibitem{Mukhanov:2003xs}
V.~F. Mukhanov, \emph{{Nucleosynthesis without a computer}},
  \href{https://doi.org/10.1023/B:IJTP.0000048169.69609.77}{\emph{Int. J.
  Theor. Phys.} {\bfseries 43} (2004) 669}
  [\href{https://arxiv.org/abs/astro-ph/0303073}{{\ttfamily
  astro-ph/0303073}}].

\bibitem{Aver:2020fon}
E.~Aver, D.~A. Berg, K.~A. Olive, R.~W. Pogge, J.~J. Salzer and E.~D. Skillman,
  \emph{{Improving helium abundance determinations with Leo P as a case
  study}}, \href{https://doi.org/10.1088/1475-7516/2021/03/027}{\emph{JCAP}
  {\bfseries 03} (2021) 027}
  [\href{https://arxiv.org/abs/2010.04180}{{\ttfamily 2010.04180}}].

\bibitem{Fernandez:2019hds}
V.~Fern\'andez, E.~Terlevich, A.~I. D\'\i{}az and R.~Terlevich, \emph{{A
  Bayesian direct method implementation to fit emission line spectra:
  Application to the primordial He abundance determination}},
  \href{https://doi.org/10.1093/mnras/stz1433}{\emph{Mon. Not. Roy. Astron.
  Soc.} {\bfseries 487} (2019) 3221}
  [\href{https://arxiv.org/abs/1905.09215}{{\ttfamily 1905.09215}}].

\bibitem{2020ApJ...896...77H}
T.~{Hsyu}, R.~J. {Cooke}, J.~X. {Prochaska} and M.~{Bolte}, \emph{{The PHLEK
  Survey: A New Determination of the Primordial Helium Abundance}},
  \href{https://doi.org/10.3847/1538-4357/ab91af}{\emph{Astrophys. J.}
  {\bfseries 896} (2020) 77}
  [\href{https://arxiv.org/abs/2005.12290}{{\ttfamily 2005.12290}}].

\bibitem{2021MNRAS.505.3624V}
M.~{Valerdi}, A.~{Peimbert} and M.~{Peimbert}, \emph{{Chemical abundances in
  seven metal-poor H II regions and a determination of the primordial helium
  abundance}}, \href{https://doi.org/10.1093/mnras/stab1543}{\emph{Mon. Not.
  Roy. Astron. Soc.} {\bfseries 505} (2021) 3624}
  [\href{https://arxiv.org/abs/2105.12260}{{\ttfamily 2105.12260}}].

\bibitem{Kurichin:2021ppm}
O.~A. Kurichin, P.~A. Kislitsyn, V.~V. Klimenko, S.~A. Balashev and A.~V.
  Ivanchik, \emph{{A new determination of the primordial helium abundance using
  the analyses of H II region spectra from SDSS}},
  \href{https://doi.org/10.1093/mnras/stab215}{\emph{Mon. Not. Roy. Astron.
  Soc.} {\bfseries 502} (2021) 3045}
  [\href{https://arxiv.org/abs/2101.09127}{{\ttfamily 2101.09127}}].

\bibitem{Cooke:2018qzw}
R.~Cooke and M.~Fumagalli, \emph{{Measurement of the primordial helium
  abundance from the intergalactic medium}},
  \href{https://doi.org/10.1038/s41550-018-0584-z}{\emph{Nature Astron.}
  {\bfseries 2} (2018) 957} [\href{https://arxiv.org/abs/1810.06561}{{\ttfamily
  1810.06561}}].

\bibitem{Planck:2018nkj}
{\scshape Planck} collaboration, N.~Aghanim et~al., \emph{{Planck 2018 results.
  I. Overview and the cosmological legacy of Planck}},
  \href{https://doi.org/10.1051/0004-6361/201833880}{\emph{Astron. Astrophys.}
  {\bfseries 641} (2020) A1}
  [\href{https://arxiv.org/abs/1807.06205}{{\ttfamily 1807.06205}}].

\bibitem{planck_legacy}
\url{https://pla.esac.esa.int/}.

\bibitem{Matsumoto:2022tlr}
A.~Matsumoto et~al., \emph{{EMPRESS. VIII. A New Determination of Primordial He
  Abundance with Extremely Metal-poor Galaxies: A Suggestion of the Lepton
  Asymmetry and Implications for the Hubble Tension}},
  \href{https://doi.org/10.3847/1538-4357/ac9ea1}{\emph{Astrophys. J.}
  {\bfseries 941} (2022) 167}
  [\href{https://arxiv.org/abs/2203.09617}{{\ttfamily 2203.09617}}].

\bibitem{Oldengott:2017tzj}
I.~M. Oldengott and D.~J. Schwarz, \emph{{Improved constraints on lepton
  asymmetry from the cosmic microwave background}},
  \href{https://doi.org/10.1209/0295-5075/119/29001}{\emph{EPL} {\bfseries 119}
  (2017) 29001} [\href{https://arxiv.org/abs/1706.01705}{{\ttfamily
  1706.01705}}].

\bibitem{Seto:2021tad}
O.~Seto and Y.~Toda, \emph{{Hubble tension in lepton asymmetric cosmology with
  an extra radiation}},
  \href{https://doi.org/10.1103/PhysRevD.104.063019}{\emph{Phys. Rev. D}
  {\bfseries 104} (2021) 063019}
  [\href{https://arxiv.org/abs/2104.04381}{{\ttfamily 2104.04381}}].

\bibitem{Kumar:2022vee}
S.~Kumar, R.~C. Nunes and P.~Yadav, \emph{{Updating non-standard neutrinos
  properties with Planck-CMB data and full-shape analysis of BOSS and eBOSS
  galaxies}},  \href{https://arxiv.org/abs/2205.04292}{{\ttfamily 2205.04292}}.

\bibitem{Burns:2022hkq}
A.-K. Burns, T.~M.~P. Tait and M.~Valli, \emph{{Indications for a Nonzero
  Lepton Asymmetry in the Early Universe}},
  \href{https://arxiv.org/abs/2206.00693}{{\ttfamily 2206.00693}}.

\bibitem{SimonsObservatory:2018koc}
{\scshape Simons Observatory} collaboration, P.~Ade et~al., \emph{{The Simons
  Observatory: Science goals and forecasts}},
  \href{https://doi.org/10.1088/1475-7516/2019/02/056}{\emph{JCAP} {\bfseries
  02} (2019) 056} [\href{https://arxiv.org/abs/1808.07445}{{\ttfamily
  1808.07445}}].

\bibitem{SimonsObservatory:2019qwx}
{\scshape Simons Observatory} collaboration, M.~H. Abitbol et~al., \emph{{The
  Simons Observatory: Astro2020 Decadal Project Whitepaper}}, {\emph{Bull. Am.
  Astron. Soc.} {\bfseries 51} (2019) 147}
  [\href{https://arxiv.org/abs/1907.08284}{{\ttfamily 1907.08284}}].

\bibitem{CMB-S4:2016ple}
{\scshape CMB-S4} collaboration, K.~N. Abazajian et~al., \emph{{CMB-S4 Science
  Book, First Edition}},  \href{https://arxiv.org/abs/1610.02743}{{\ttfamily
  1610.02743}}.

\bibitem{Abazajian:2019eic}
K.~Abazajian et~al., \emph{{CMB-S4 Science Case, Reference Design, and Project
  Plan}},  \href{https://arxiv.org/abs/1907.04473}{{\ttfamily 1907.04473}}.

\bibitem{EscuderoAbenza:2020cmq}
M.~Escudero~Abenza, \emph{{Precision early universe thermodynamics made simple:
  $N_{\rm eff}$ and neutrino decoupling in the Standard Model and beyond}},
  \href{https://doi.org/10.1088/1475-7516/2020/05/048}{\emph{JCAP} {\bfseries
  05} (2020) 048} [\href{https://arxiv.org/abs/2001.04466}{{\ttfamily
  2001.04466}}].

\bibitem{Fields:2019pfx}
B.~D. Fields, K.~A. Olive, T.-H. Yeh and C.~Young, \emph{{Big-Bang
  Nucleosynthesis after Planck}},
  \href{https://doi.org/10.1088/1475-7516/2020/03/010}{\emph{JCAP} {\bfseries
  03} (2020) 010} [\href{https://arxiv.org/abs/1912.01132}{{\ttfamily
  1912.01132}}].

\bibitem{Akita:2020szl}
K.~Akita and M.~Yamaguchi, \emph{{A precision calculation of relic neutrino
  decoupling}},
  \href{https://doi.org/10.1088/1475-7516/2020/08/012}{\emph{JCAP} {\bfseries
  08} (2020) 012} [\href{https://arxiv.org/abs/2005.07047}{{\ttfamily
  2005.07047}}].

\bibitem{Froustey:2020mcq}
J.~Froustey, C.~Pitrou and M.~C. Volpe, \emph{{Neutrino decoupling including
  flavour oscillations and primordial nucleosynthesis}},
  \href{https://doi.org/10.1088/1475-7516/2020/12/015}{\emph{JCAP} {\bfseries
  12} (2020) 015} [\href{https://arxiv.org/abs/2008.01074}{{\ttfamily
  2008.01074}}].

\bibitem{Bennett:2020zkv}
J.~J. Bennett, G.~Buldgen, P.~F. De~Salas, M.~Drewes, S.~Gariazzo, S.~Pastor
  et~al., \emph{{Towards a precision calculation of $N_{\rm eff}$ in the
  Standard Model II: Neutrino decoupling in the presence of flavour
  oscillations and finite-temperature QED}},
  \href{https://doi.org/10.1088/1475-7516/2021/04/073}{\emph{JCAP} {\bfseries
  04} (2021) 073} [\href{https://arxiv.org/abs/2012.02726}{{\ttfamily
  2012.02726}}].

\bibitem{Dolgov:2002ab}
A.~D. Dolgov, S.~H. Hansen, S.~Pastor, S.~T. Petcov, G.~G. Raffelt and D.~V.
  Semikoz, \emph{{Cosmological bounds on neutrino degeneracy improved by flavor
  oscillations}},
  \href{https://doi.org/10.1016/S0550-3213(02)00274-2}{\emph{Nucl. Phys. B}
  {\bfseries 632} (2002) 363}
  [\href{https://arxiv.org/abs/hep-ph/0201287}{{\ttfamily hep-ph/0201287}}].

\bibitem{Wong:2002fa}
Y.~Y.~Y. Wong, \emph{{Analytical treatment of neutrino asymmetry equilibration
  from flavor oscillations in the early universe}},
  \href{https://doi.org/10.1103/PhysRevD.66.025015}{\emph{Phys. Rev. D}
  {\bfseries 66} (2002) 025015}
  [\href{https://arxiv.org/abs/hep-ph/0203180}{{\ttfamily hep-ph/0203180}}].

\bibitem{Abazajian:2002qx}
K.~N. Abazajian, J.~F. Beacom and N.~F. Bell, \emph{{Stringent Constraints on
  Cosmological Neutrino Antineutrino Asymmetries from Synchronized Flavor
  Transformation}},
  \href{https://doi.org/10.1103/PhysRevD.66.013008}{\emph{Phys. Rev. D}
  {\bfseries 66} (2002) 013008}
  [\href{https://arxiv.org/abs/astro-ph/0203442}{{\ttfamily
  astro-ph/0203442}}].

\bibitem{Froustey:2021azz}
J.~Froustey and C.~Pitrou, \emph{{Primordial neutrino asymmetry evolution with
  full mean-field effects and collisions}},
  \href{https://doi.org/10.1088/1475-7516/2022/03/065}{\emph{JCAP} {\bfseries
  03} (2022) 065} [\href{https://arxiv.org/abs/2110.11889}{{\ttfamily
  2110.11889}}].

\bibitem{pdg}
{\scshape Particle Data Group} collaboration, R.~L. Workman and Others,
  \emph{{Review of Particle Physics}},
  \href{https://doi.org/10.1093/ptep/ptac097}{\emph{PTEP} {\bfseries 2022}
  (2022) 083C01}.

\bibitem{Cooke:2017cwo}
R.~J. Cooke, M.~Pettini and C.~C. Steidel, \emph{{One Percent Determination of
  the Primordial Deuterium Abundance}},
  \href{https://doi.org/10.3847/1538-4357/aaab53}{\emph{Astrophys. J.}
  {\bfseries 855} (2018) 102}
  [\href{https://arxiv.org/abs/1710.11129}{{\ttfamily 1710.11129}}].

\bibitem{Pisanti:2007hk}
O.~Pisanti, A.~Cirillo, S.~Esposito, F.~Iocco, G.~Mangano, G.~Miele et~al.,
  \emph{{PArthENoPE: Public Algorithm Evaluating the Nucleosynthesis of
  Primordial Elements}},
  \href{https://doi.org/10.1016/j.cpc.2008.02.015}{\emph{Comput. Phys. Commun.}
  {\bfseries 178} (2008) 956}
  [\href{https://arxiv.org/abs/0705.0290}{{\ttfamily 0705.0290}}].

\bibitem{Consiglio:2017pot}
R.~Consiglio, P.~F. de~Salas, G.~Mangano, G.~Miele, S.~Pastor and O.~Pisanti,
  \emph{{PArthENoPE reloaded}},
  \href{https://doi.org/10.1016/j.cpc.2018.06.022}{\emph{Comput. Phys. Commun.}
  {\bfseries 233} (2018) 237}
  [\href{https://arxiv.org/abs/1712.04378}{{\ttfamily 1712.04378}}].

\bibitem{Gariazzo:2021iiu}
S.~Gariazzo, P.~F.~de Salas, O.~Pisanti and R.~Consiglio, \emph{{PArthENoPE
  revolutions}}, \href{https://doi.org/10.1016/j.cpc.2021.108205}{\emph{Comput.
  Phys. Commun.} {\bfseries 271} (2022) 108205}
  [\href{https://arxiv.org/abs/2103.05027}{{\ttfamily 2103.05027}}].

\bibitem{deSalas:2015glj}
P.~F. de~Salas, M.~Lattanzi, G.~Mangano, G.~Miele, S.~Pastor and O.~Pisanti,
  \emph{{Bounds on very low reheating scenarios after Planck}},
  \href{https://doi.org/10.1103/PhysRevD.92.123534}{\emph{Phys. Rev. D}
  {\bfseries 92} (2015) 123534}
  [\href{https://arxiv.org/abs/1511.00672}{{\ttfamily 1511.00672}}].

\bibitem{Hasegawa:2019jsa}
T.~Hasegawa, N.~Hiroshima, K.~Kohri, R.~S.~L. Hansen, T.~Tram and S.~Hannestad,
  \emph{{MeV-scale reheating temperature and thermalization of oscillating
  neutrinos by radiative and hadronic decays of massive particles}},
  \href{https://doi.org/10.1088/1475-7516/2019/12/012}{\emph{JCAP} {\bfseries
  12} (2019) 012} [\href{https://arxiv.org/abs/1908.10189}{{\ttfamily
  1908.10189}}].

\bibitem{Nollett:2013pwa}
K.~M. Nollett and G.~Steigman, \emph{{BBN And The CMB Constrain Light,
  Electromagnetically Coupled WIMPs}},
  \href{https://doi.org/10.1103/PhysRevD.89.083508}{\emph{Phys. Rev. D}
  {\bfseries 89} (2014) 083508}
  [\href{https://arxiv.org/abs/1312.5725}{{\ttfamily 1312.5725}}].

\bibitem{Sabti:2019mhn}
N.~Sabti, J.~Alvey, M.~Escudero, M.~Fairbairn and D.~Blas, \emph{{Refined
  Bounds on MeV-scale Thermal Dark Sectors from BBN and the CMB}},
  \href{https://doi.org/10.1088/1475-7516/2020/01/004}{\emph{JCAP} {\bfseries
  01} (2020) 004} [\href{https://arxiv.org/abs/1910.01649}{{\ttfamily
  1910.01649}}].

\bibitem{Ichikawa:2005vw}
K.~Ichikawa, M.~Kawasaki and F.~Takahashi, \emph{{The Oscillation effects on
  thermalization of the neutrinos in the Universe with low reheating
  temperature}}, \href{https://doi.org/10.1103/PhysRevD.72.043522}{\emph{Phys.
  Rev. D} {\bfseries 72} (2005) 043522}
  [\href{https://arxiv.org/abs/astro-ph/0505395}{{\ttfamily
  astro-ph/0505395}}].

\bibitem{Kolb:1986nf}
E.~W. Kolb, M.~S. Turner and T.~P. Walker, \emph{{The Effect of Interacting
  Particles on Primordial Nucleosynthesis}},
  \href{https://doi.org/10.1103/PhysRevD.34.2197}{\emph{Phys. Rev. D}
  {\bfseries 34} (1986) 2197}.

\bibitem{Grohs:2019cae}
E.~B. Grohs, J.~R. Bond, R.~J. Cooke, G.~M. Fuller, J.~Meyers and M.~W. Paris,
  \emph{{Big Bang Nucleosynthesis and Neutrino Cosmology}},
  \href{https://arxiv.org/abs/1903.09187}{{\ttfamily 1903.09187}}.

\bibitem{Pitrou:2021vqr}
C.~Pitrou, A.~Coc, J.-P. Uzan and E.~Vangioni, \emph{{Resolving conclusions
  about the early Universe requires accurate nuclear measurements}},
  \href{https://doi.org/10.1038/s42254-021-00294-6}{\emph{Nature Rev. Phys.}
  {\bfseries 3} (2021) 231} [\href{https://arxiv.org/abs/2104.11148}{{\ttfamily
  2104.11148}}].

\end{thebibliography}\endgroup

\end{document}